\documentclass{article}

\usepackage{algorithm}
\usepackage{algorithmic}
\usepackage{microtype}
\usepackage{graphicx}
\usepackage{subfigure}
\usepackage{booktabs} 
\usepackage{amsmath}
\usepackage{amssymb}
\usepackage{amsthm}
\usepackage{xspace}

\DeclareRobustCommand{\Sec}[1]{Sec.~\ref{sec:#1}}

\DeclareRobustCommand{\App}[1]{App.~\ref{app:#1}}

\DeclareRobustCommand{\Fig}[1]{Fig.~\ref{fig:#1}}

\DeclareRobustCommand{\Eq}[1]{Eq.~(\ref{eq:#1})}

\DeclareRobustCommand{\Alg}[1]{Algorithm~\ref{alg:#1}}

\newcommand{\eqn}[1]{\begin{align}#1\end{align}}
\allowdisplaybreaks

\newcommand{\auq}[1]{\textnormal{AUQ}\left[{#1}\right]}
\newcommand{\e}[1]{\mathbb{E}\left[{#1}\right]}
\newcommand{\var}[1]{\mathbb{V}\left[{#1}\right]}
\DeclareMathOperator*{\argmax}{arg\,max}

\renewcommand{\algorithmicrequire}{\textbf{Input:}}
\renewcommand{\algorithmicensure}{\textbf{Output:}}

\newcommand{\Stroke}{{\sc Stroke}\xspace}
\newcommand{\CriteoVisit}{{\sc CriteoVisit}\xspace}
\newcommand{\CriteoConversion}{{\sc CriteoConversion}\xspace}
\newcommand{\RetailHero}{{\sc RetailHero}\xspace}

\PassOptionsToPackage{hyphens}{url}\usepackage{hyperref}

\usepackage[accepted]{StyleGuide2023/icml2023}

\icmltitlerunning{Task-specific experimental design for treatment effect estimation}

\usepackage[skip=0pt]{caption} 

\begin{document}

\twocolumn[
\icmltitle{Task-specific experimental design for treatment effect estimation}



\icmlsetsymbol{equal}{*}

\begin{icmlauthorlist}
\icmlauthor{Bethany Connolly}{equal}
\icmlauthor{~~~Kim Moore}{equal}
\icmlauthor{~~~Tobias Schwedes}{}
\icmlauthor{~~~Alexander Adam}{} \\ \vskip 1mm
\icmlauthor{Gary Willis}{}
\icmlauthor{~~~Ilya Feige}{}
\icmlauthor{~~~Christopher Frye}{fac}
\end{icmlauthorlist}

\icmlaffiliation{fac}{Faculty, 160 Old Street, London, UK}
\icmlcorrespondingauthor{Christopher Frye}{chris.f@faculty.ai}


\vskip 0.3in
]



\printAffiliationsAndNotice{\icmlEqualContribution} 

\begin{abstract}
%
Understanding causality should be a core requirement of any attempt to build real impact through AI.
%
Due to the inherent unobservability of counterfactuals, large randomised trials (RCTs) are the standard for causal inference.
%
But large experiments are generically expensive, and randomisation carries its own costs, e.g.~when suboptimal decisions are trialed.
%
Recent work has proposed more sample-efficient alternatives to RCTs, but these are not adaptable to the downstream application for which the causal effect is sought.
%
In this work, we develop a task-specific approach to experimental design and derive sampling strategies customised to particular downstream applications.
%
Across a range of important tasks, real-world datasets, and sample sizes, our method outperforms other benchmarks, e.g.~requiring an order-of-magnitude less data to match RCT performance on targeted marketing tasks.
\end{abstract}

\vspace{-4mm}

\section{Introduction}

Artificial intelligence makes its impact through the influence it has on actions taken in the real world. This influence is often indirect, e.g.~when machine learning predictions inform human decision making. The field of causal inference studies the effect of actions on outcomes directly. Causal inference is thus central to the broader programme of engineering AI's positive effects on society.

Treatment effect estimation is a paradigm of causal inference that ranges from measuring the average treatment effect (ATE) of an intervention at the population level to the more granular prediction of individual treatment effects (ITE). Such treatment efficacy measures are useful across domains, from education \cite{olaya2020uplift} and customer retention \cite{devriendt2021you} to medicine \cite{qi2019predicting} and marketing \cite{verbeke2023or}.

The fundamental challenge of causal inference is the inaccessibility of counterfactuals: upon observing an individual's response to one treatment, their response had they received a different treatment is not observable. Treatment effect estimation must therefore occur through observation of many individuals, and large randomised controlled trials are the hallmark of experimental design \cite{devriendt2018literature}. RCTs minimise bias in cohort selection and treatment assignment to produce treatment groups representative of the population at large \cite{markozannes2021survey}. 

At smaller sample sizes, however, randomised treatment assignments often result in feature imbalances between treatment groups \cite{morgan2012rerandomization}. This has spurred substantial research into covariate balancing between treatment groups \cite{greevy2004optimal, morgan2012rerandomization, kallus2018optimal}. These techniques focus on treatment assignment given a fixed cohort of trial participants.

There is recent interest in optimising cohort selection as well \mbox{\cite{qin2021budgeted, jesson2021causal, addanki2022sample}.} Such work aims to reduce the sample size required for performant treatment effect estimation. This is fundamentally an active learning (AL)  \cite{ren2021survey} task wherein a population of unknown treatment response represents a pool of unlabelled data; a point is labelled when that individual's outcome is observed in a trial \mbox{\cite{puha2020batch}.} In this context, AL can further be viewed as a multi-armed bandit problem \cite{deng2011active}.

In our view, AL techniques designed for supervised learning are ill-suited for causal modelling. Due to the fundamental challenge of causal inference, treatment effect models must be trained on loss functions that do not directly target the treatment effect. The \mbox{T-learner} \cite{gutierrez2017causal}, for example, is the difference in two models trained separately to predict outcomes in each treatment group. Such proxy tasks are often far removed from the downstream application, e.g.~ordering subjects by ITE. Loss-targeting AL methods \cite{roy2001toward} are thus ill-fit for causal problems. Uncertainty-based AL \cite{beluch2018power} is not well-suited either: in treatment effect estimation, one aims to learn the \emph{difference} in two outcome probabilities, irrespective of their entropies. 

In this work, we remedy this problem by introducing a new approach to sample selection for treatment effect estimation. We take inspiration from \citet{mindermann2022prioritized}, focusing on data points that are ``worth learning,'' though we do not employ the RHO loss presented in that work as it too is inherently supervised. Instead we derive a series of analytic sample-selection strategies tailored to specific applications.

\subsection{Our contributions}

In this paper, we are concerned with the optimisation of \emph{experimental design}, i.e.~the process of sampling individuals from a large population and assigning them treatments for a trial. The outcomes that result are used to model treatment effects and these predictions are then used to perform some downstream task. Our primary contributions include:
\begin{enumerate}
\item We introduce a novel approach to experimental design specific to the downstream task to be performed.
\item As an important example, we present an algorithm for experimental design for cases in which the area under the Qini curve (AUQ) \cite{radcliffe2011real} is a good proxy for performance on the downstream task.
\item We validate our approach on several large real-world datasets and demonstrate state-of-the-art sample efficiency. In particular, on AUQ tasks our method requires about an order-of-magnitude less data to match the performance achieved by an RCT.
\end{enumerate}
Our empirical analysis on {large} {real-world} datasets is itself significant, as tests on small semi-synthetic datasets are the norm in much of the related literature. We show that our method (i) scales well to large populations, (ii) leads unequivocally to statistically significant increases in performance, and (iii) works well on real-world data, whereas semi-synthetic data is often unrealistically simplistic.

\section{Background}

Let $\mathcal{X}$ denote a domain of features with density $p(X)$, and let $\mathcal{T}=\{0,1\}$ represent a binary set of treatments that can be applied to individuals in the domain. Experimental design refers to the process of selecting individuals from $\mathcal X$, assigning treatments from $\mathcal T$, and observing outcomes that take values in $\mathcal{Y}$. In particular, we are interested in learning treatment effects. Provided no unobserved variables confound treatment and outcome, the individual treatment effect, or \emph{uplift}, for $x\in \mathcal{X}$ is given by
\begin{align}\label{eq:uplift}
    u(x) = \e{Y|X=x, T=1} - \e{Y|X=x, T=0}
\end{align}
We use $\hat u(x)$ to denote a learnt estimate of the uplift. The \emph{average treatment effect} can be written as:
\begin{align}\label{eq:ate}
    \textnormal{ATE}[u] ~&= \int_{\mathcal{X}} p(x) \, u(x) \, dx
\end{align}

\subsection{Metrics}
\label{sec:metrics}

A variety of metrics exist to judge a learnt treatment effect, each being the primary measure of performance in a specific context. In practice,  one estimates these metrics using an \emph{RCT test set}, in which individuals are sampled from $p(X)$ and their treatments $T$ are uniformly randomised.

\subsubsection*{Mean squared error} 

First consider the \emph{mean squared error} (MSE) of the model:
\begin{equation}\label{eq:mse}
    \text{MSE}[\hat u]\, = \int_\mathcal{X} p(x) \, (\hat{u}(x) - u(x))^2 \, dx
\end{equation}
In this work, the MSE principally serves as a simple example to demonstrate our method. The MSE is used primarily in the context of synthetic data where $u(x)$ is known exactly. The PEHE (precision in estimation of heterogeneous effects) is a common proxy for MSE in practice \mbox{\cite{qin2021budgeted}.}

\subsubsection*{Squared ATE error} 

In a clinical context, the purpose of a trial is often to estimate the ATE of a particular intervention on a population. The \emph{squared ATE error} $\big( \text{ATE}[\hat u] - \text{ATE}[u] \big)^2$ is therefore a useful target metric for this context. The true ATE$[u]$ can be estimated well as a simple difference in empirical means $\e{Y|T=1} - \e{Y|T=0}$ on an RCT test set.

\subsubsection*{Qini curve and AUQ}

In the context of targeted marketing, the task is often to determine which subset of one's customer base should receive a finite budget of advertisements to maximise return on investment. A model $\hat u(x)$ can perform this task by selecting those customers with highest predicted uplift. The \emph{Qini curve} $Q(f)$ measures performance at this task by tracing out the expected increase in per-capita outcomes as a function of the fraction $f$ of customers accommodated by the budget \cite{radcliffe2007using}. The Qini curve is maximised as the model $\hat u(x)$ tends to the ground truth $u(x)$.

To write down an estimator of the Qini curve on a finite sample $\{x_1, \ldots, x_n\}$ from the underlying population, one must first order this sample according to descending predicted uplift: let $\{\hat x_1, \ldots \hat x_n\}$ denote the ordering in which $\hat u(\hat x_i) \geq \hat u(\hat x_{i+1})$ for all $i$. Then
\begin{equation}
Q(f) \approx \frac{1}{n} \, \sum_{i=1}^{f \cdot n} \, u(\hat x_i)
\end{equation}
Note this estimator references the true uplift distribution; \Alg{qini} in the appendix details the procedure to evaluate the Qini curve on an RCT test set, in which $u(x)$ is of course unobserved. The \emph{area under the Qini curve} provides a general metric on model performance across budgets:
\begin{equation}\label{eq:auq}
    \auq{\hat{u}} \,=\, \int_0^1 df \, Q(f) - \frac{\text{ATE}[u]}{2}
\end{equation}
We subtract off the area under the straight line between $(0, 0)$ and $(1, \text{ATE}[u])$ as this is the Qini curve of a random baseline model \cite{gutierrez2017causal}.

In \App{erupt}, we describe one additional metric known as ERUPT (expected response under prescribed treatments).

\section{Theory}
\label{sec:our_method}

Here we develop an approach to experimental design that optimises performance on a downstream task, as quantified by a metric $\mathcal R$. This target metric generally differs from the loss function used to train any model.

First, we make a simplifying assumption, taking a discrete feature space $\mathcal X = \{x_1, x_2, \ldots, x_k\}$. This will not limit the applicability of our method: we present an approach to discretisation in \Sec{discrete} that allows us to apply our method on continuous data. Also, for concreteness, we assume binary outcomes $\mathcal Y = \{0, 1\}$; footnote \ref{foot:cont} describes how to alter our results for continuous $\mathcal Y$. 

Second, we aim to optimise $\mathbb E[ \mathcal R[\hat U]]$, and we need to select a representative model $\hat U$ to insert into this expression. We opt to work with \Eq{bayesian_model} below as it allows us to derive a sampling strategy by hand. This choice does not limit the applicability of our method either: any downstream model can be trained on the sample that results from our method, and in \Sec{exp} we empirically demonstrate excellent performance for a variety of model types.

To motivate our choice of $\hat U$, note that we would have complete knowledge of the treatment effect if we had access to $p(Y=1 | X=x, T=t) =: \theta_{xt}^*$. Let us use $\theta_{xt}$ to denote any prediction of these values (and we’ll suppress subscripts when referring to the full set $\{\theta_{xt}\}$). Then, given a prior $p_\alpha(\Theta=\theta)$ with hyperparameter dependence indicated by $\alpha$, and given observations $\{(x_i, t_i, y_i)\}_{i=1}^n$, there exists a posterior predictive $p_\alpha(Y\,|\,X=x,\,T=t,\,\{(x_i, t_i, y_i)\}_{i=1}^n)$. Motivated by \Eq{uplift}, we take the difference in the means of the posterior predictives for $T = 1$ and $0$ as our representative uplift model, and we place a Beta distribution prior over $\theta_{xt}$, with hyperparameters\footnote{To accommodate the case of continuous outcomes, one could substitute $\alpha \to \mu\,\nu$ and $\beta \to \nu\,(1 - \mu)$ in \Eq{bayesian_model} where $\mu$ and $\nu$ are the parameters of a normal-gamma prior.\label{foot:cont}} $\alpha_{xt}$ and $\beta_{xt}$, as it is conjugate to the Bernoulli likelihood. This results in
\eqn{
\hat U(x) \,~&=~ \sum_{t = 0}^1 \, (-1)^{t+1} \: \frac{\alpha_{xt} ~+~ \sum_{i=1}^{n(x, t)} \,Y_{i(x, t)}}{\alpha_{xt} ~+~ \beta_{xt} ~+~ n(x, t)} \label{eq:bayesian_model}
}
In this equation, $n(x, t)$ denotes the number of samples in $\{(x_i, t_i, y_i)\}_{i=1}^n$ with $x_i=x$ and $t_i=t$, and we have capitalised $\hat U(x)$ and $Y_{i(x,t)}$ to emphasise that the model is a random variable through its dependence on outcomes $Y_{i(x, t)} \sim p(Y|X=x,T=t)$. The target metric $\mathcal R[\hat U]$ is a random variable as well, and our method involves calculating its expectation value by hand:
\eqn{
\mathbb E\big[ \mathcal R[\hat U] \big] ~=:~ f_\alpha^{\mathcal R}\big(\{n(x, t)\} \,;\, \theta^* \big)
}
The function $f_\alpha^{\mathcal R}\big(\{n(x, t)\} \,;\, \theta^* \big)$ is computed explicitly for a few key target metrics in the following section.

Now, $\theta^*$ is unknown in practice, and $\mathcal R[\hat U]$ is generically unobserved during the experiment as well. Nonetheless, the latent value of the target metric $\mathcal R[\hat U]$ will change at each step $n(x, t) \mapsto n(x, t) + 1$, as this amounts to drawing an outcome $Y_{n(x,t)+1} \sim p(Y|X=x, T=t)$ and using it to update the model $\hat U$. Each improvement in $\mathcal R[\hat U]$ can be viewed as a reward. The true reward distribution belongs to a family of distributions indexed by $\theta$, each with expectation value $f_\alpha^{\mathcal R}\big(\{n(x, t)\} \,;\, \theta \big)$, and every observation tightens the posterior $p_\alpha\big(\Theta=\theta \,\big|\, \{(x_i, t_i, y_i)\}_{i=1}^{n}\big)$.

We thus have a $2k$-armed bandit wherein pulling the $(x, t)$ arm accrues a reward and improves the estimation of future rewards. This system can be optimised through Thompson sampling \cite{thompson1933likelihood}, wherein at each step one draws a value of $\theta$ from the posterior, then selects the next $(x, t)$ that maximises the expected reward given $\theta$.

We therefore propose an approach to experimental design in which one cycles through the following sequence:
\eqn{
&\text{1. Draw:} ~~ \theta \sim p_\alpha\big(\Theta=\theta \,\big|\, \{(x_i, t_i, y_i)\}_{i=1}^n\big) \nonumber\\[3pt]
&\text{2. Select:} ~~\argmax_{(x^*, t^*)} ~~ f_\alpha^{\mathcal R}\!\left(\left\{\!\!\!
{\footnotesize\begin{array}{lc}
n(x^*, t^*) + 1 & \text{at } (x^*, t^*) \\
n(x, t) & \text{otherwise}
\end{array}}
\!\!\!\right\}; \,\theta\! \right) \nonumber\\[1pt]
&\text{3. Observe:} ~~ Y_{n(x^*,t^*)+1} \sim p(Y|X=x^*, T=t^*) \nonumber\\[7pt]
&\text{4. Update:} ~~ p_\alpha\big(\Theta=\theta \,\big|\, \{(x_i, t_i, y_i)\}_{i=1}^{n+1}\big) \nonumber
}
Thompson sampling optimises $\mathbb E[ \mathcal R[\hat U]]$ in two ways: it \mbox{results} in an asymptotically (i.e. as $n \to \infty$) optimal sampling policy, and it achieves minimal regret, i.e.~rewards forgone due to suboptimality at finite $n$ are minimised \cite{agrawal2012analysis,kaufmann2012thompson}. 

Our full method, including discretisation (see \Sec{discrete}), is detailed in \Alg{method} and summarised in \Fig{schematic}. It can be run in a sequential/online fashion as described here, if observations can be made between selection steps. Otherwise, it should be run in batches by iterating between steps 1-2 and deferring steps 3-4 to be performed at longer intervals; see \App{addl_exp} and \Fig{batch}.

\subsection{Calculations for example target metrics}
\label{sec:calculations}

Here we provide examples of $f_\alpha^{\mathcal R}\big(\{n(x, t)\} \,;\, \theta^* \big)$ for a few key target metrics. The bias and variance of our representative model, \Eq{bayesian_model}, arise throughout these calculations, so we collect results for these quantities here. (See \App{extra_theory} for all derivations of the results in this section.)
\eqn{
\mathbb E[\hat U(x) - u(x)] \,~&=~ \sum_{t=0}^1 \: (-1)^{t+1} \, \frac{\alpha_{xt} - (\alpha_{xt} + \beta_{xt}) \, \theta^*_{xt}}{\alpha_{xt} + \beta_{xt} + n(x, t)} \nonumber \\
\mathbb V[\hat U(x)] \,~&=~ \sum_{t=0}^1 \, \frac{\theta^*_{xt}\,(1-\theta^*_{xt})\,n(x, t)}{\big(\alpha_{xt} + \beta_{xt} + n(x, t)\big)^2}  \label{eq:bias_var}
}
In these equations, we replaced \mbox{$\mathbb E[Y|X=x, T=t]$} and $\mathbb V[Y|X=x, T=t]$  with $\theta^*_{xt}$ and $\theta^*_{xt}\,(1-\theta^*_{xt})$ respectively, as $Y$ is Bernoulli distributed.

\subsubsection*{Mean squared error}

Beginning with the simplest case, consider the MSE, which takes the following expected value on $\hat U$:
\eqn{
\mathbb E\big[ \text{MSE}[\hat U] \big] 
&= \sum_{x\in \mathcal{X}} \, p(x) ~ \mathbb E\big[(\hat{U}(x) - u(x))^2\big] \label{eq:f_mse} \\[3pt]
&= \sum_{x\in\mathcal{X}} \, p(x) \, \Big(\mathbb E\big[\hat U(x) - u(x)\big]^2 + \mathbb V\big[\hat U(x)\big] \Big) \nonumber \\[4pt]
&=: -f_\alpha^\text{MSE}\big(\{n(x, t)\};\,\theta^*\big) \nonumber
}
where one can refer to \Eq{bias_var} to write down the explicit dependence of $f_\alpha^\text{MSE}(\{n(x, t)\};\,\theta^*)$ on its arguments.
We refer to the process of using \Eq{f_mse} to iterate through our 4-step method of \Sec{our_method} as \emph{MSE-optimised sampling}.

\subsubsection*{Squared ATE error}

Similar to the case of the MSE, the squared ATE error takes the following expected value on $\hat U$:
\eqn{
    &f_\alpha^\text{ATE}\big(\{n(x, t)\};\,\theta^*\big) := \label{eq:f_ate} \\[5pt]
    &-\Big(\sum_{x \in \mathcal X} \, p(x) \, \mathbb E\big[ \hat U(x) - u(x) \big] \Big)^2 - \sum_{x \in \mathcal X} \, p(x)^2 \, \mathbb V\big[\hat U(x)\big] \nonumber
}
The explicit dependence of $f_\alpha^\text{ATE}(\{n(x, t)\};\,\theta^*)$ on its arguments follows again from \Eq{bias_var}. \emph{ATE-optimised sampling} refers to the insertion of \Eq{f_ate} into the method of \Sec{our_method}.

\subsubsection*{Area under Qini curve}

The AUQ is more complicated but in many contexts much more relevant \cite{radcliffe2011real, rzepakowski2012decision, gutierrez2017causal}. In the large $n$ limit, $\mathbb E[\text{AUQ}[\hat U]]$ takes the following value:
\eqn{
&f_\alpha^\text{AUQ}\big(\{n(x, t)\};\,\theta^*\big) := \label{eq:f_auq} \\[2pt] 
&\sum_{\substack{x, x'\in\mathcal{X}\\ x\ne x'}} p(x) \, p(x') \, u(x')\left[\frac{1}{2} + \frac{1}{2}\,\textnormal{erf}\left(\frac{u(x') - u(x)}{\sqrt{2 \, \sigma^2(x, x')}}\right)\right] \nonumber
}
where the dependence of $f_\alpha^\text{AUQ}\big(\{n(x, t)\};\,\theta^*\big)$ on its arguments follows from $u(x) = \theta^*_{x1} - \theta^*_{x0}$ and the shorthand
\eqn{
\sigma^2(x,x') = \sum_{t=0}^1 \, \frac{\theta^*_{xt}\,(1 - \theta^*_{xt})}{n(x,t)} + (x \to x')
}
We refer to the insertion of \Eq{f_auq} into the method of \Sec{our_method} as \emph{AUQ-optimised sampling}.

In \App{erupt_deriv}, we derive $f_\alpha^{\mathcal R}\big(\{n(x, t)\} \,;\, \theta^* \big)$ for the ERUPT metric as an additional example.

\subsection{Intuition for task-specific sampling}
\label{sec:intuition}

\begin{figure}[t!]
\begin{center}
\centerline{\includegraphics[width=\columnwidth]{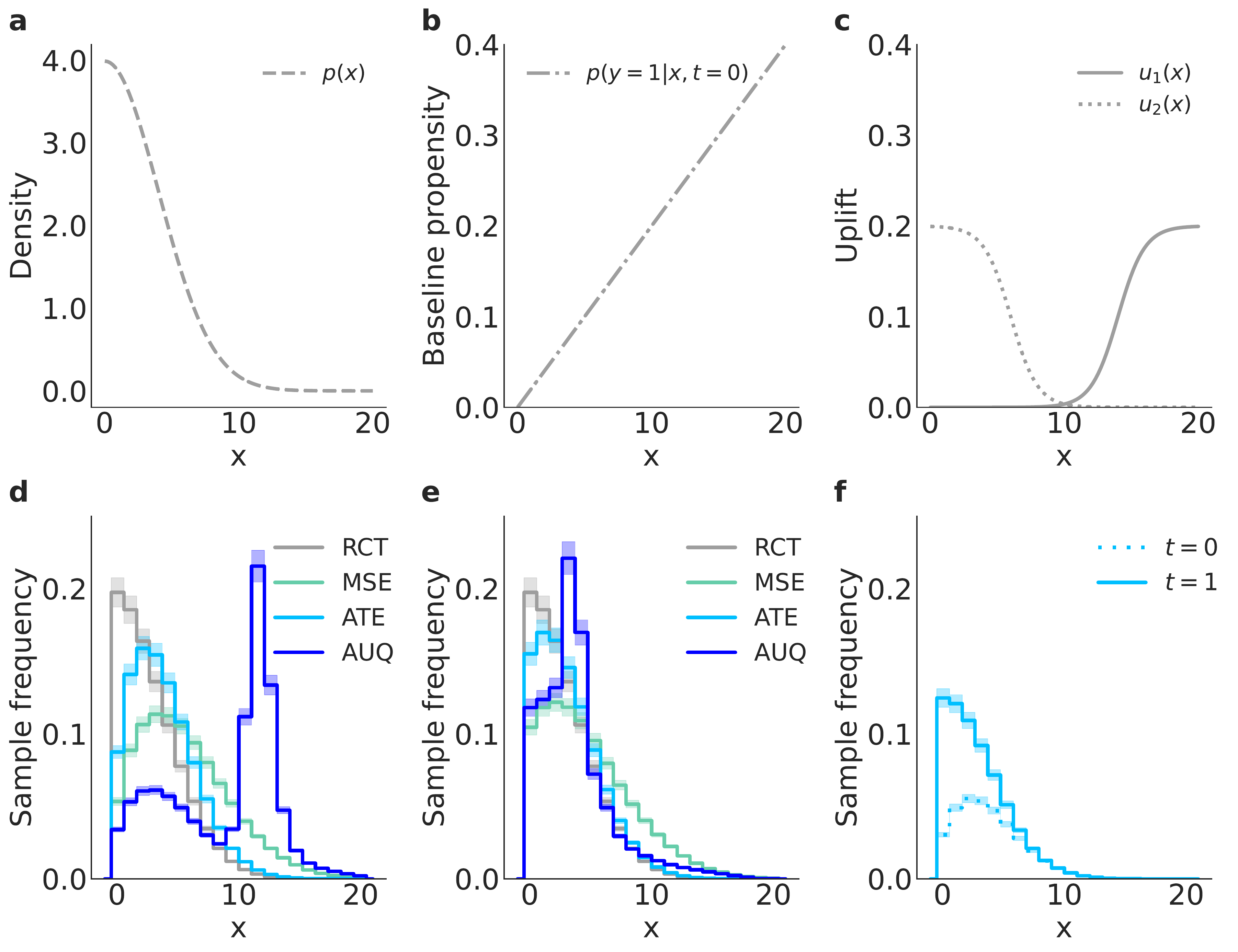}}
\vskip 1mm
\caption{Synthetic density (a), baseline propensity (b), and uplift (c) distributions, along with the samples that result from our optimised methods; (d) and (e) correspond to $u_1$ and $u_2$, respectively, and (f) shows the ATE-optimised sample from (e) split by treatment and control.
}
\label{fig:intuition}
\end{center}
\vskip -6mm
\end{figure}

Here we demonstrate our optimised samplers on synthetic data. (See \App{experimental} for details of this experiment.)

\begin{figure*}[t!]
\begin{center}
\centerline{\includegraphics[width=0.67\textwidth]{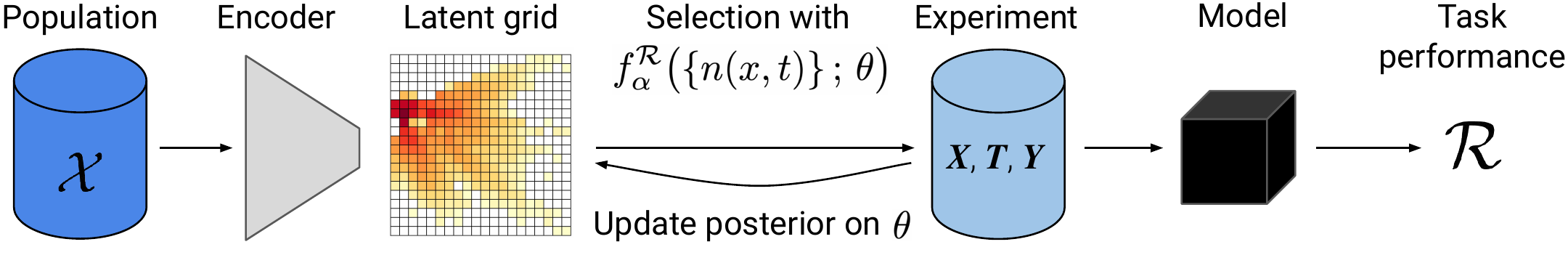}}
\caption{Schematic of our end-to-end method for task-specific experimental design for treatment effect estimation.}
\label{fig:schematic}
\end{center}
\vskip -5mm
\end{figure*}

For this experiment, we take $\mathcal X = \{1, 2, \ldots, 20\}$ with a truncated Gaussian density $p(X)$ as shown in \Fig{intuition}(a). We set the \emph{baseline propensity}, i.e.~$p(Y=1\,|\,X=x,\,T=0)$, to a linear function of $x$ as depicted in \Fig{intuition}(b). We consider two uplift distributions, $u_1(x)$ and $u_2(x)$, mirror images of each other defined using sigmoids; see \Fig{intuition}(c). 

In this section, we aim to provide intuition for the different $f_\alpha^{\mathcal R}(\{n(x, t\}; \theta^*)$'s derived in \Sec{calculations}. To isolate the effect of this function, we cycle through the 4-step algorithm of \Sec{our_method} but with $p_\alpha\big(\Theta=\theta \,\big|\, \{(x_i, t_i, y_i)\}_{i=1}^{n}\big)$ set to a point mass at the correct value of $\theta^*$. We set $\alpha_{xt} = \beta_{xt} = 0$ in $f_\alpha^{\mathcal R}(\{n(x, t\}; \theta^*)$ for concreteness as well.

The resulting distributions are plotted in \Fig{intuition}(d) for $u_1(x)$ and \Fig{intuition}(e) for $u_2(x)$, where they are compared against an RCT sample. \Fig{intuition}(f) shows the ATE-optimised sample from \Fig{intuition}(e) split into its $T=0$ and $1$ components. Our samplers generically select different distributions for treatment and control to optimise the target metric.

Note from \Fig{intuition} that the RCT sample is proportional to the density, whereas the MSE-optimised sample is also sensitive to the variance in outcomes (see Eq.~\ref{eq:f_mse}) which pulls the distribution to the right. The ATE-optimised sample comes out roughly in between, as it has stronger (i.e.~squared) dependence on the density (see Eq.~\ref{eq:f_ate}). The AUQ-optimised sample is additionally sensitive to uplift differences $u(x) - u(x')$ (see Eq.~\ref{eq:f_auq}) which leads to a strong peak coincident with the sharp gradient in $u(x)$.

\subsection{Discretisation for general datatypes}
\label{sec:discrete}

Our method, as developed so far, assumes a discrete feature space. Here we describe our approach to discretisation to accommodate continuous and mixed-type data.

We discretise data using a variational autoencoder (VAE) \cite{kingma2013auto} trained on $\mathcal X$. In particular, we encode the dataset and place a rectangular grid on latent space to partition $\mathcal X$ into a finite set $\{x_1, x_2, \ldots, x_k\}$ as required. This is the final ingredient in our approach to experimental design; see \Alg{method} and \Fig{schematic}.

This approach to discretisation partitions the data in a smooth homogeneous space rather than the heterogeneous space of raw features. It also gives us control over the cardinality $k$, which influences the runtime of our algorithm. We thus find it useful to perform latent-space discretisation regardless of the datatypes in $\mathcal X$. \App{extra_theory} shows further that discretisation does not introduce bias into our method.

\begin{algorithm}[t!]
\caption{Task-specific experimental design}\label{alg:method}
\begin{algorithmic}{
\vskip 7pt
\REQUIRE 
Population $\mathcal{X}$\,;~
Target metric $\mathcal R$\,;~
Sample size $n$ \\[7pt]
VAE's latent representation of population $\mathcal{X}_\text{latent}$ \\
Rectangular grid on latent space $\mathcal{B}$ \\
Prior parameters $\alpha_{bt}$ and $\beta_{bt}$ for each $b\in\mathcal{B}$ and $t\in \mathcal{T}$ \\[7pt]
\ENSURE 
Sample of experimental results $\{(x_i, t_i, y_i)\}_{i=1}^n$ \\[7pt]
\STATE Initialise $n_{bt}=0$ for $b\in\mathcal{B}$ and $t\in\mathcal{T}$ \\[1pt]
\FOR{$i=1$ \TO $n$} 
\STATE Draw $\theta_{bt} \sim \text{Beta}(\alpha_{bt}, \beta_{bt})$ for each $b\in\mathcal{B}$ and $t\in \mathcal{T}$ \\[1pt]
\FOR{each $b \in \mathcal B$ and $t \in \mathcal T$} 
\STATE Set $n'_{b't'} = n_{b't'}$ for each $b' \in \mathcal B$ and $t' \in \mathcal T$ \\[1pt]
\STATE Increment $n'_{bt} \gets n'_{bt} + 1$ \\[1pt]
\STATE Set $f_{bt} = f_\alpha^{\mathcal R}(\{n'_{b't'}\}; \theta)$ using $\mathcal R$-specific formula \\[1pt]
\ENDFOR \\[1pt]
\STATE Select $(b_i, t_i) = \argmax_{(b, t)} f_{bt}$ \\[1pt]
\STATE Sample $x_\text{latent}$ uniformly from $\mathcal X_\text{latent} \cap b_i$ \\[1pt]
\STATE Select individual $x_i \in \mathcal X$ that corresponds to $x_\text{latent}$ \\[1pt]
\STATE Observe $y_i \sim p(Y | X=x_i, T=t_i)$ \\[1pt]
\STATE $n_{b_i t_i} \gets n_{b_i t_i} + 1$  \\[1pt]
\STATE $\alpha_{b_i t_i} \gets \alpha_{b_i t_i} + y_i$  \\[1pt]
\STATE $\beta_{b_i t_i} \gets \beta_{b_i t_i} + (1 - y_i)$ \\[3pt]
\ENDFOR
\RETURN $\{(x_i, t_i, y_i)\}_{i=1}^n$
}\end{algorithmic}
\end{algorithm}

\subsubsection*{Uniform sampling in latent space}

Having defined a grid on latent space, one simple but novel alternative to \Alg{method} is to select grid cells and assign treatments both uniformly at random.  We refer to this as \emph{uniform sampling in latent space}. As it is task-agnostic by construction, one cannot expect this strategy to perform optimally across metrics. Indeed, \Fig{ate_uniform} in the appendix shows poor performance for the ATE task. Despite this, we show in \Sec{real} that uniform sampling in latent space is remarkably effective for the AUQ task. \App{importance} develops an importance-sampling formalism that could be used to determine when uniform sampling will perform well.

In \App{addl_exp}, we isolate the advantage of discretisation with a VAE by comparing uniform sampling in latent space to uniform sampling in \emph{feature} space; see \Fig{feature-space} in the appendix.

\section{Experiments}
\label{sec:exp}

Here we present empirical analyses of our approach to experimental design. See \App{experimental} for full experimental details.

\subsection{Validation on synthetic data}

\begin{figure}[t!]
\begin{center}
\centerline{\includegraphics[width=\columnwidth, trim={0 0 0 2cm}, clip]{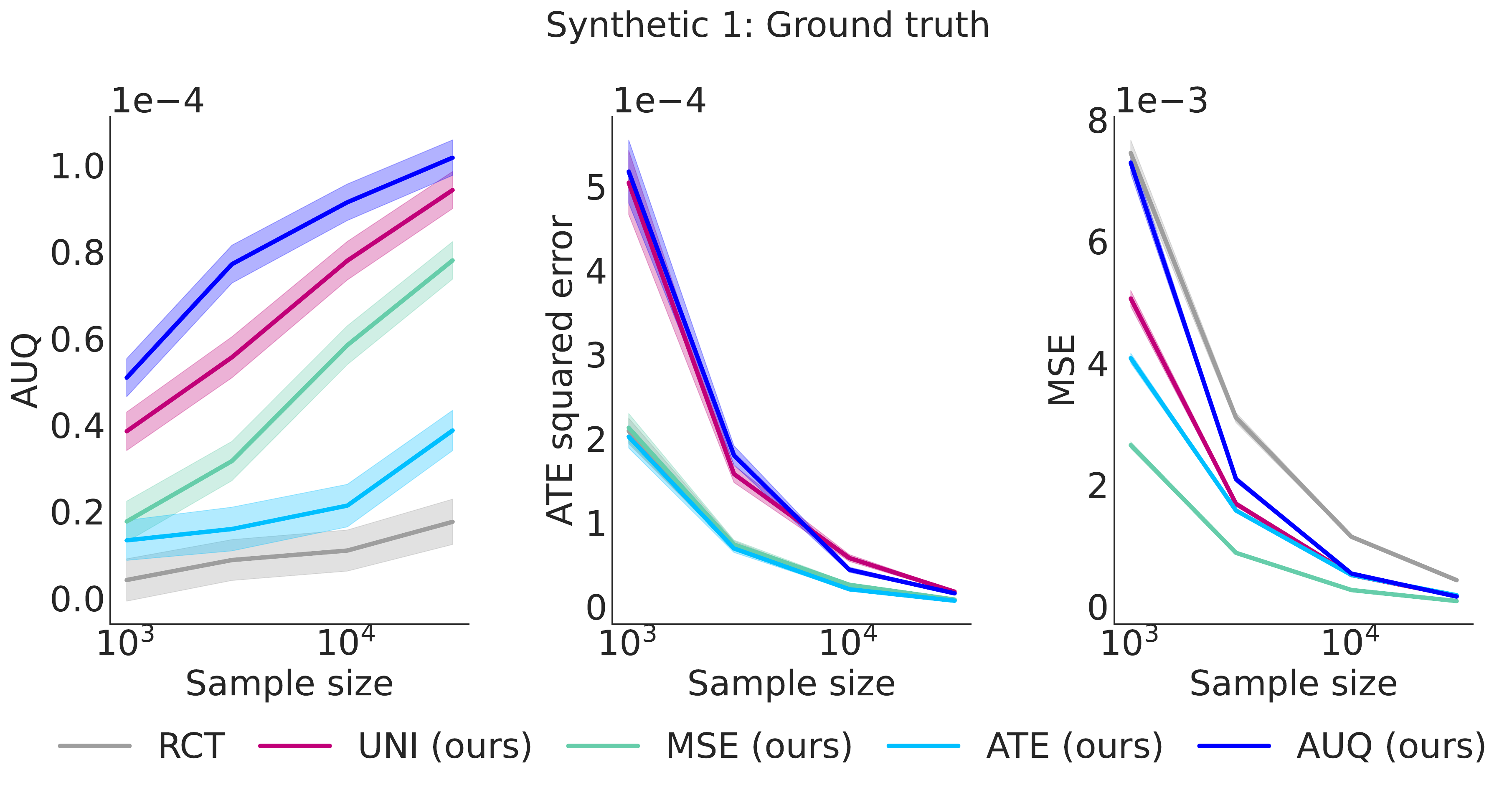}}
\vskip 1mm
\caption{Performance of MSE-, ATE-, and AUQ-optimised samplers across target metrics on synthetic data, compared to RCT and uniform sampling on $\mathcal X$.}
\label{fig:synthetic_performance}
\end{center}
\vskip -8mm
\end{figure}

In \Sec{intuition} we used two synthetic datasets to show distributions resulting from our optimised samplers. Here we take the $u_1(x)$ case, use the samples from \Fig{intuition}(d) to train uplift models according to \Eq{bayesian_model}, and measure model performance according to the metrics of \Sec{metrics}. \Fig{synthetic_performance} displays the results, with one plot for each target metric and one curve for each optimised sampler. We also include ordinary RCT sampling as well as uniform sampling in latent space for comparison. Note that, across sample sizes, the MSE\,/\,ATE\,/\,AUQ\,-\,optimised sampler outperforms the alternatives when judged by the test-set MSE\,/\,ATE\,/\,AUQ, respectively. This provides the first empirical validation of our approach to task-specific experimental design.

\subsection{Performance on real-world data}
\label{sec:real}

Here we describe experiments carried out on real-world data, but first we address a subtlety in our setup. We aim to benchmark a new method of experimental design using data gathered from historical experiments that were performed as RCTs. We circumvent this apparent paradox as follows. Most of the datasets we use are very large in size; in particular, after discretisation there are many examples of most $(x, t)$ combinations we would be interested in sampling in our experimental design. We thus treat the large pool of data as though treatments were yet unassigned, and whenever the dataset lacks an $(x, t)$ that our algorithm selects, we select the algorithm's next preference instead.

The datasets we use for our experiments are described at length in \App{datasets}. In brief, we test our method on:
\vspace{-1mm}
\begin{itemize}
\item \Stroke: clinical trial evaluating aspirin's effect on stroke patients; our sub-selection procedure results in a dataset of size 9k \cite{sandercock2011international}.
\vspace{-1mm}
\item \CriteoVisit \& \CriteoConversion: marketing trial evaluating effectiveness of email campaign on two different outcomes; we sub-select 7M rows of data \cite{Diemert2018}.
\vspace{-1mm}
\item \RetailHero: marketing trial in which we engineered features from purchase history data for 200k individuals (see \App{datasets} for references).
\end{itemize}

Next we describe performance results for two downstream applications: clinical trials and targeted marketing.

\begin{figure}[t!]
\begin{center}
\centerline{\includegraphics[width=0.7\columnwidth, clip]{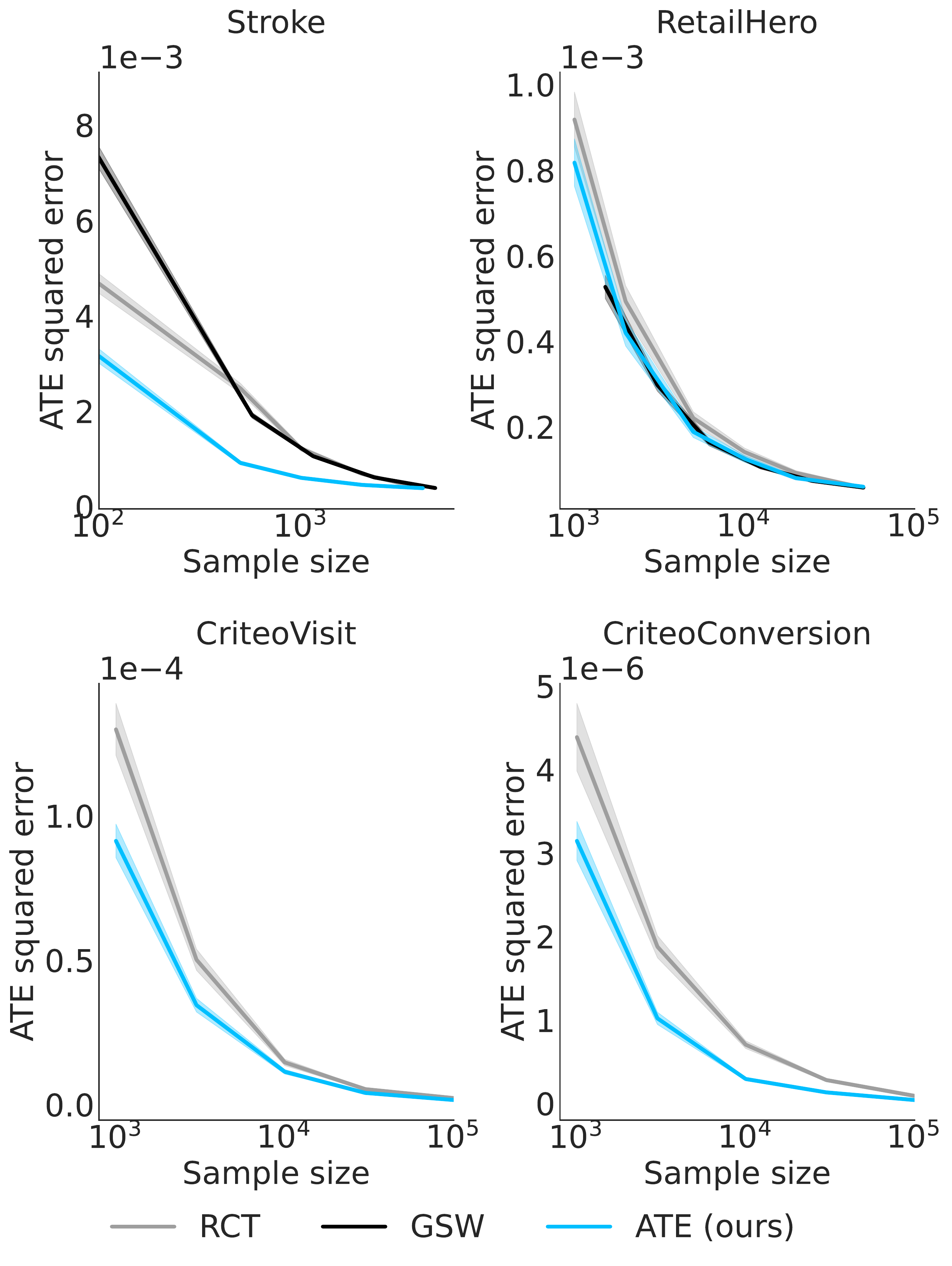}}
\caption{Performance of ATE-optimised sampler across datasets, benchmarked against RCT and Recursive GSW.}
\label{fig:ate-estimation}
\end{center}
\vskip -8mm
\end{figure}

\subsubsection*{Application to clinical trials}

Here we take the squared ATE error as our target metric. While \Stroke is our only example of a \emph{true} clinical trial, we include all our datasets in this study. In this experiment, we sample data according to different algorithms, then reweight the sample according to the population density to estimate the ATE.

In \Fig{ate-estimation} we show the performance of our ATE-optimised sampler alongside RCT. On three of four datasets, our ATE-optimised sampler provides a meaningful advantage over RCT with respect to the target metric. 

We also benchmark the Recursive Gram-Schmidt Walk (GSW) method of \citet{addanki2022sample} though it does not outperform RCT on \Stroke or \RetailHero. Due to the quadratic scaling of the GSW algorithm with respect to population size \cite{harshaw2019balancing}, Recursive GSW is infeasible on both {\sc Criteo} datasets. \Fig{ate_uniform} in the appendix further benchmarks the B-EMCMITE method of \citet{puha2020batch}. This is a loss-targeting method of supervised AL applied to
causal modelling, which is only computationally feasible for very small populations. We find its ATE performance to be poor in our experiments.

\Fig{time-complexity} in the appendix benchmarks the time complexity of our method against Recursive GSW and B-EMCMITE as a function of population size. This is a very favourable result for our method, as the only step in \Alg{method} that scales with population size is the uniform sampling of an occupant from the selected latent grid cell.

\subsubsection*{Application to targeted marketing}

Here we take the AUQ as our target metric and exclude \Stroke from this study, as the AUQ requires a very large test set to estimate precisely. In this experiment, we sample data according to different algorithms, then train T-learners \cite{gutierrez2017causal} on the samples to predict uplift, and finally measure the AUQ of each T-learner.

\begin{figure}[t!]
\begin{center}
\centerline{\includegraphics[width=1.04\columnwidth]{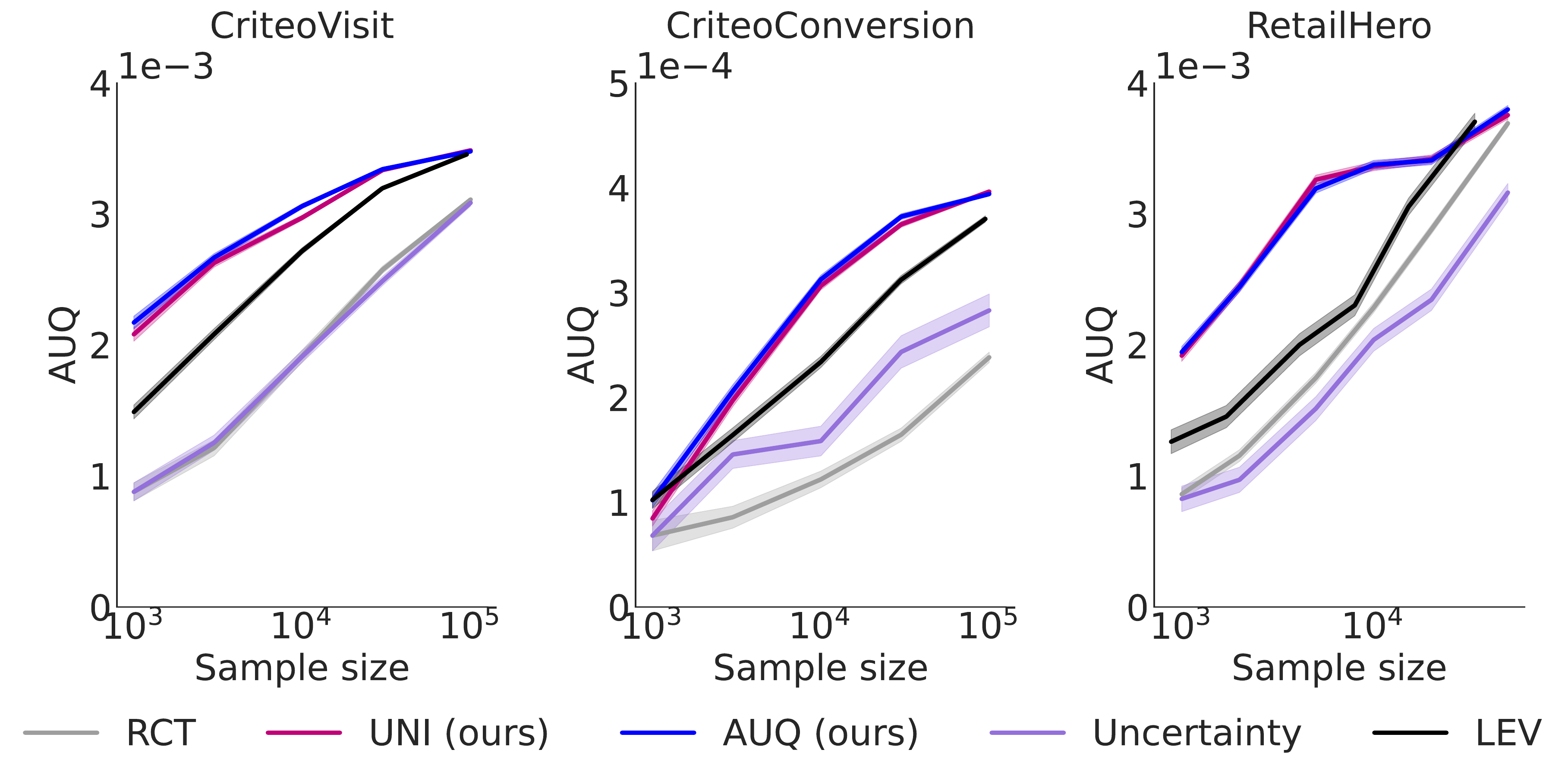}}
\caption{Performance of AUQ-optimised sampling and uniform sampling in latent space across datasets, benchmarked against RCT, Leverage Scoring, and uncertainty sampling.}
\label{fig:marketing}
\end{center}
\vskip -7mm
\end{figure}

In \Fig{marketing} we show the performance of our AUQ-optimised sampler. Across all sample sizes and datasets, our method provides a substantial improvement over RCT, generally matching its performance with roughly an order-of-magnitude smaller sample. 

\Fig{marketing} also benchmarks the Leverage Scoring method of \citet{addanki2022sample}, which performs between our method and RCT, as well as an uncertainty-based approach to supervised AL \cite{beluch2018power} in which points are selected
that have the highest entropy in predicted outcomes, according to the model-in-training. The latter is not tailored to causal modelling and performs commensurate with RCT.

\Fig{marketing} further shows that uniform sampling in latent space performs on par with AUQ-optimised sampling. This is quite remarkable, given the state-of-the-art nature of our AUQ-optimised sampler and the simplicity of our uniform sampler. We understand this as follows:

AUQ-optimised sampling targets segments of the population where uplift differences are large; this is a direct result of \Eq{f_auq} that we demonstrated in \Sec{intuition}. In addition, our sampler has the largest uncertainty on uplift in regions that have been under-sampled. Through our Thompson sampling procedure of \Sec{our_method}, our sampler will attribute the largest swings in uplift -- and thus the largest uplift differences -- to the least-sampled regions of the data. This can result in a sampling strategy that is not too different from uniform. (See \Fig{rct-uni-auq-latentsamples} in the appendix for an example).

\App{addl_exp} includes two further results that we mention here: (i) Whereas \Fig{marketing} displays the mean AUQ of the various samplers, \Fig{box_whisker} shows the distribution of AUQ across trials to give a sense of how often our method outperforms RCT in practice. (ii) \Fig{erupt} explores an additional use case, ERUPT-optimised sampling, where our method outperforms RCT, Leverage Scoring, and uniform sampling in latent space.

\subsection{Sensitivity to hyperparameters}
\label{sec:hyperparams}

Here we evaluate the sensitivity of our method to the hyperparameters required to fully define it.

\subsubsection*{Latent space dimensionality}

\begin{figure}[t!]
\begin{center}
\centerline{\includegraphics[width=\columnwidth]{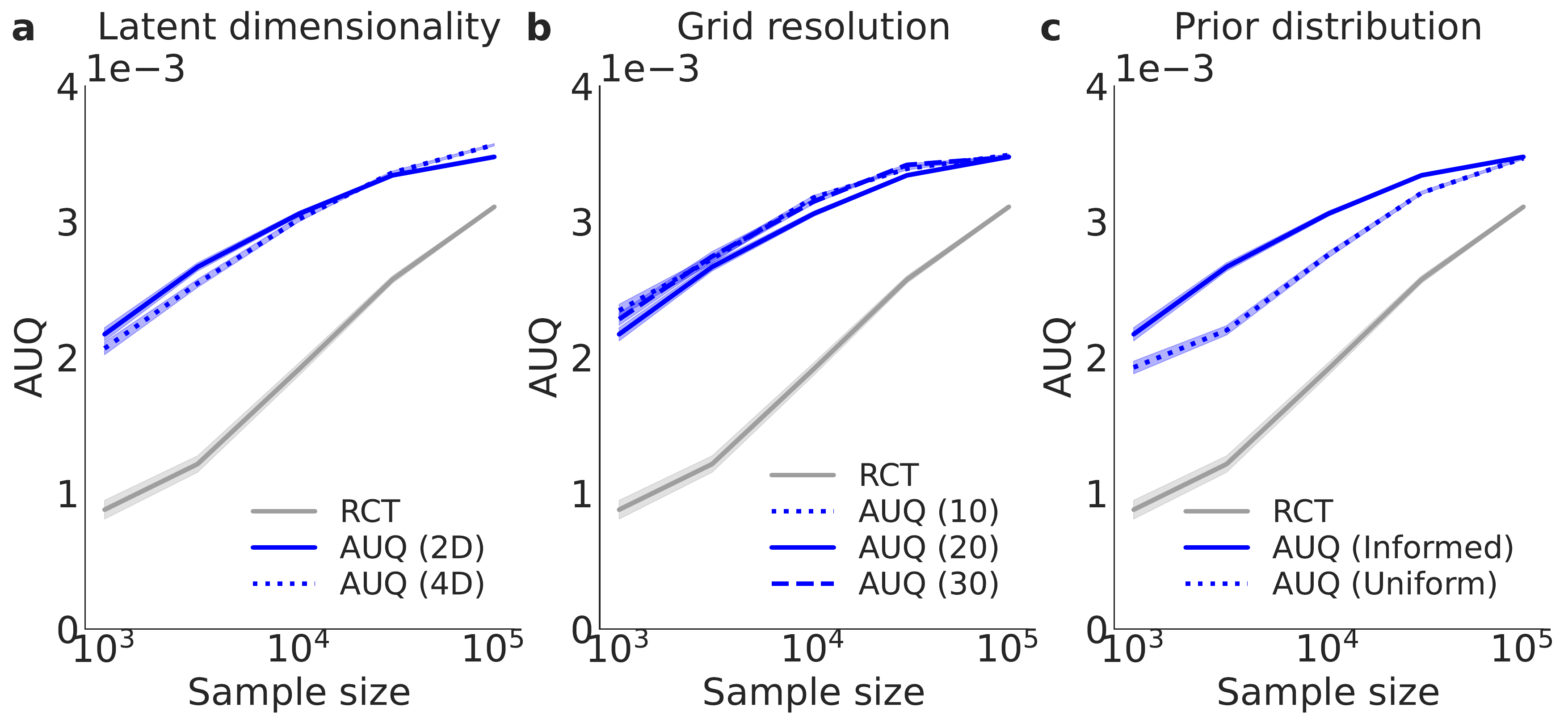}}
\vskip 2mm
\caption{Hyperparameter sensitivity of AUQ-optimised sampling to latent dimension, grid resolution, and prior choice on \CriteoVisit, with RCT as an anchor point.}
\label{fig:dim-epsilon-prior}
\end{center}
\vskip -5mm
\end{figure}

Provided latent space has capacity to capture the factors of variation underlying the raw data, we would not expect strong dependence on the choice of latent dimensionality. Our experiments above were performed with latent dimension 2, and in \Fig{dim-epsilon-prior}(a) we test our AUQ-optimised sampler on \CriteoVisit with latent dimension 4. The result is that performance hardly varies with this hyperparameter. (Though note that latent dimensionality would need to be tuned more carefully for higher-dimensional feature spaces.)

\subsubsection*{Sampling grid resolution}

Regarding the rectangular grid used to discretise latent space, one would expect a trade-off: a finer grid allows more precise targeting, while a coarser grid allows lower-variance $\theta_{xt}$ estimates (see \Sec{our_method}). Our experiments above all used 20x20 grids, and in \Fig{dim-epsilon-prior}(b) we test our AUQ-optimised sampler on \CriteoVisit with 10x10 and 30x30 grids, showing minimal dependence on this hyperparameter.

\subsubsection*{Prior distribution}

The prior (in particular $\alpha_{xt}$ and $\beta_{xt}$) directly influences the formula used to select the training sample (see \Sec{our_method}) so we should expect dependence on this choice. 

For the ATE experiments above, we used $\alpha_{xt}=\beta_{xt}=1$, placing uniform priors on the Bernoulli outcome probabilities. For the AUQ experiments, we defaulted to an informed prior with $\alpha_{xt} = 1 + p(x) \, n^{+}_{t}$ and $\beta_{xt} = 1 + p(x) \, n_{t}^{-}$. Here $n_{t}^\pm$ is a ballpark estimate of the successes/failures anticipated for treatment group $t$ (but agnostic of $x$) in an experiment of the given size. Thus $\alpha_{xt}$ and $\beta_{xt}$ reflect the anticipated base rates in the data. (These estimates need not be accurate; \Fig{informed_prior_sensitivity} in the appendix shows minimal effect when the $n^+_{t}$'s are doubled or halved.) 

In \Fig{dim-epsilon-prior}(c) we show how AUQ-optimised sampling on \CriteoVisit depends on the prior. The informed prior does provide a meaningful advantage over the uniform prior when it comes to AUQ performance, though even the uninformed prior provides a large improvement over RCT.

\subsubsection*{Uplift model type}

Finally we demonstrate that our method works well for a variety of downstream uplift models. While we derived our algorithm  using one specific model (Eq.~\ref{eq:bayesian_model}) our AUQ experiments used a T-learner \cite{gutierrez2017causal} for the modelling task. The good performance of \Fig{marketing} thus provides initial validation of this claim. 

\begin{figure}[t!]
\begin{center}
\centerline{\includegraphics[width=\columnwidth]{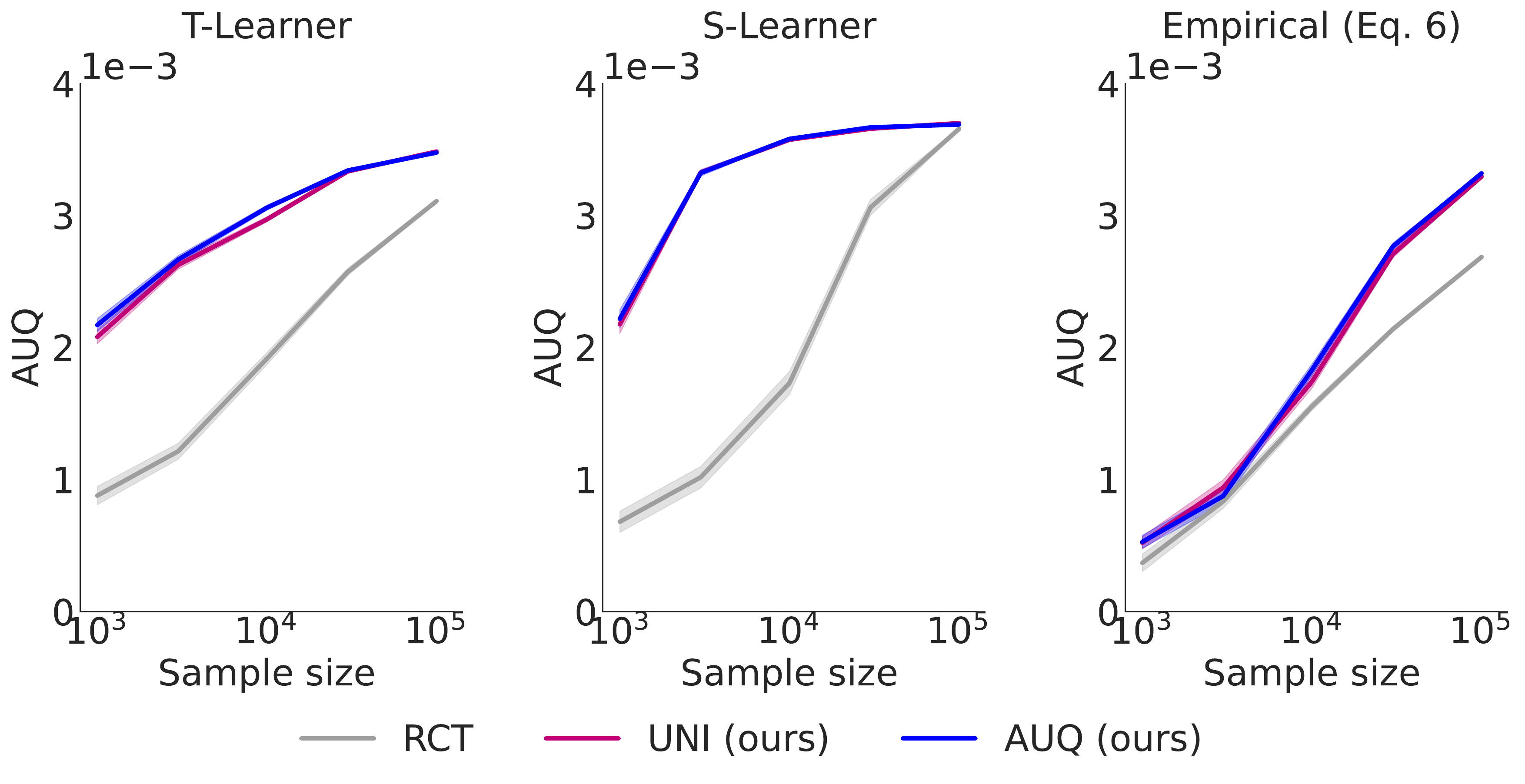}}
\caption{Variation of the downstream model trained on AUQ-optimised, uniform, and RCT samples from \CriteoVisit. Our methods outperform RCT across model choices.}
\label{fig:models}
\end{center}
\vskip -9mm
\end{figure}

In \Fig{models} we validate this further by showing the T-learner's performance trained on AUQ-optimised samples from \CriteoVisit alongside that of an S-learner and of \Eq{bayesian_model} itself (with $\alpha_{xt}=\beta_{xt}=0$). While absolute performance varies across model types, we find that training a model on a sample selected with our method dramatically outperforms that same model trained on an RCT sample.

\subsubsection*{Target metric \& batch size}

\App{addl_exp} includes two further results that we mention here for completeness. First, \Fig{task-specificity} shows the performance of our AUQ-, ATE-, and MSE-optimised samplers evaluated on both the AUQ and ATE-squared-error metrics on \CriteoVisit. This demonstrates that no sampler, not even RCT, performs well across target metrics as disparate as AUQ and ATE squared error. This result calls into question any attempt at task-agnostic experimental design and provides further justification for our general approach.

Second, \Fig{batch} shows that our AUQ-optimised sampler is extremely insensitive to the batch size with which observed outcomes are used to update its posterior (cf.~step 4 in \Sec{our_method} and the discussion below it). The ATE-optimised sampler is more sensitive but still demonstrates useful performance.

\section{Related work}
\label{sec:relatedwork}

Here we compare our work to other studies of experimental design for treatment effect estimation.

\citet{deng2011active} frame experimental design as a multi-armed bandit problem as we do. They arrive at a formula similar to our $f_\alpha^\text{MSE}(\{n(x, t)\};\theta)$ with $\alpha_{xt}=\beta_{xt}=0$, but leaving out dependence on the density $p(x)$. In comparison, we consider our approach to be more general as we consider a variety of target metrics, including the AUQ which we have not seen targeted in any other experimental design work. \citet{deng2011active} also assume the data is split into subpopulations but do not include any algorithm to do so -- a hurdle to applying their method in practice.

\citet{puha2020batch} devise an approach to experimental design called B-EMCMITE, which targets individuals who maximise the expected change in the uplift model's loss function, taking inspiration from supervised AL. This is downstream-task-agnostic, in contrast with our approach. We benchmark our ATE-optimised sampler against B-EMCMITE in \Fig{ate_uniform}, at least where computationally feasible, and find that our method strongly outperforms it. B-EMCMITE scales poorly as it requires model gradients across the entire population at every selection step; see \Fig{time-complexity}.
 
\citet{addanki2022sample} present two sampling algorithms, Recursive GSW and Leverage Scoring, intended for ATE and ITE estimation, respectively. They prove a notion of optimality under very restrictive assumptions -- e.g.~that outcomes, and thus the uplift itself, are linear functions of the features. In \Sec{real} we show that our optimised samplers outperform Recursive GSW and Leverage Scoring across tasks, datasets, and sample sizes -- except for the ATE task on \RetailHero (see \Fig{ate-estimation}) where results are even.

Our work relates to pool-based AL with expected-model-change querying \cite{ren2021survey}. The defining distinction is that we do not query according to the uplift model's loss function; instead, our approach targets the metric that defines performance on the downstream task.

Our method of uniform sampling in latent space relates to core-set methods \cite{sener2018active} in AL. VAEs have been used elsewhere to obtain low-dimensional representations for general AL \cite{pourkamali2019effectiveness, sinha2019variational}, though we have not seen this applied to the problem of causal modelling outside our work.

\section{Conclusion}

We have introduced a novel method of experimental design for treatment effect learning, in which a task-specific formula governs cohort selection and treatment assignment. We derived this formula explicitly for four metrics -- MSE, ATE, AUQ, ERUPT -- which find relevance in distinct applications. Our method outperforms RCT and other benchmarks almost universally across the tasks, datasets, and sample sizes we studied. Most notably, on the AUQ metric prevalent in contexts in which the ITE is of primary interest, our method requires about an order-of-magnitude less data to match RCT performance. 

In our view, the primary limitation of our method is that it requires by-hand calculations for each target metric, and future work could explore data-driven alternatives to this requirement. Follow-up work could also explore alternatives to Thompson sampling as the backbone of our method. More generally, we hope our work spurs further advances in the field, so that efficient causal inference can have a positive impact across a wide range of important applications.



\bibliography{bibliography}
\bibliographystyle{StyleGuide2023/icml2023}

\appendix

\section{Additional experimental results}
\label{app:addl_exp}

In this appendix, we provide the additional experimental results that were referenced throughout the paper. 

\begin{figure}[t!]
\begin{center}
\centerline{\includegraphics[width=\columnwidth]{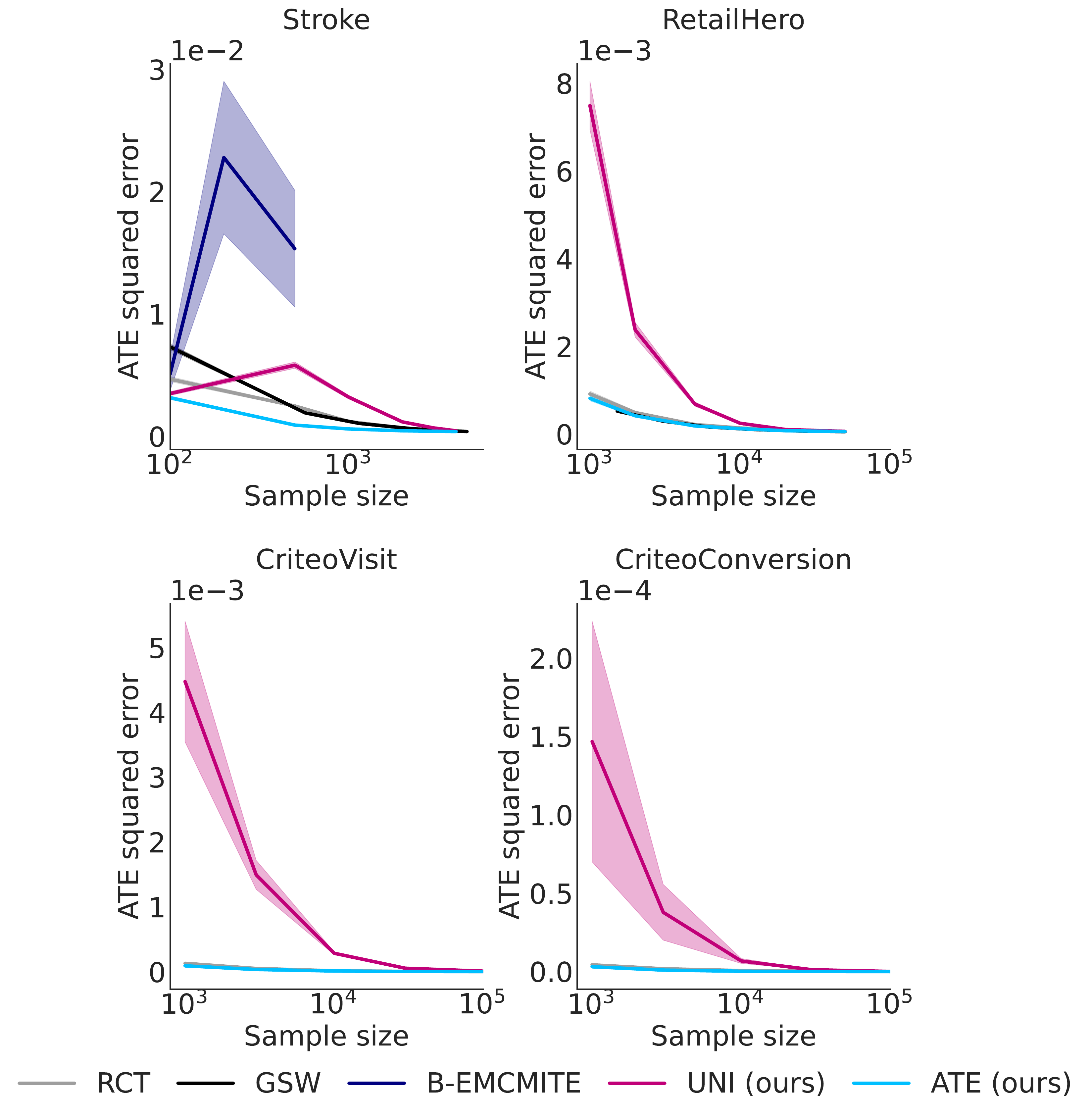}}
\caption{Performance of ATE-optimised sampler across datasets benchmarked against RCT, uniform sampling in latent space, Recursive GSW, and B-EMCMITE.}
\label{fig:ate_uniform}
\vskip -5mm
\end{center}
\end{figure}

\Fig{ate_uniform} displays the performance of ATE-optimised sampling and uniform sampling in latent space across datasets, benchmarked against RCT, Recursive GSW \cite{addanki2022sample} and B-EMCMITE \cite{puha2020batch}. Across datasets, this result shows that uniform sampling in latent space is not well-aligned with the ATE prediction task. It also shows that B-EMCMITE, where computationally feasible, is quite misaligned as well.

\begin{figure}[t!]
\begin{center}
\centerline{\includegraphics[width=0.82\columnwidth]{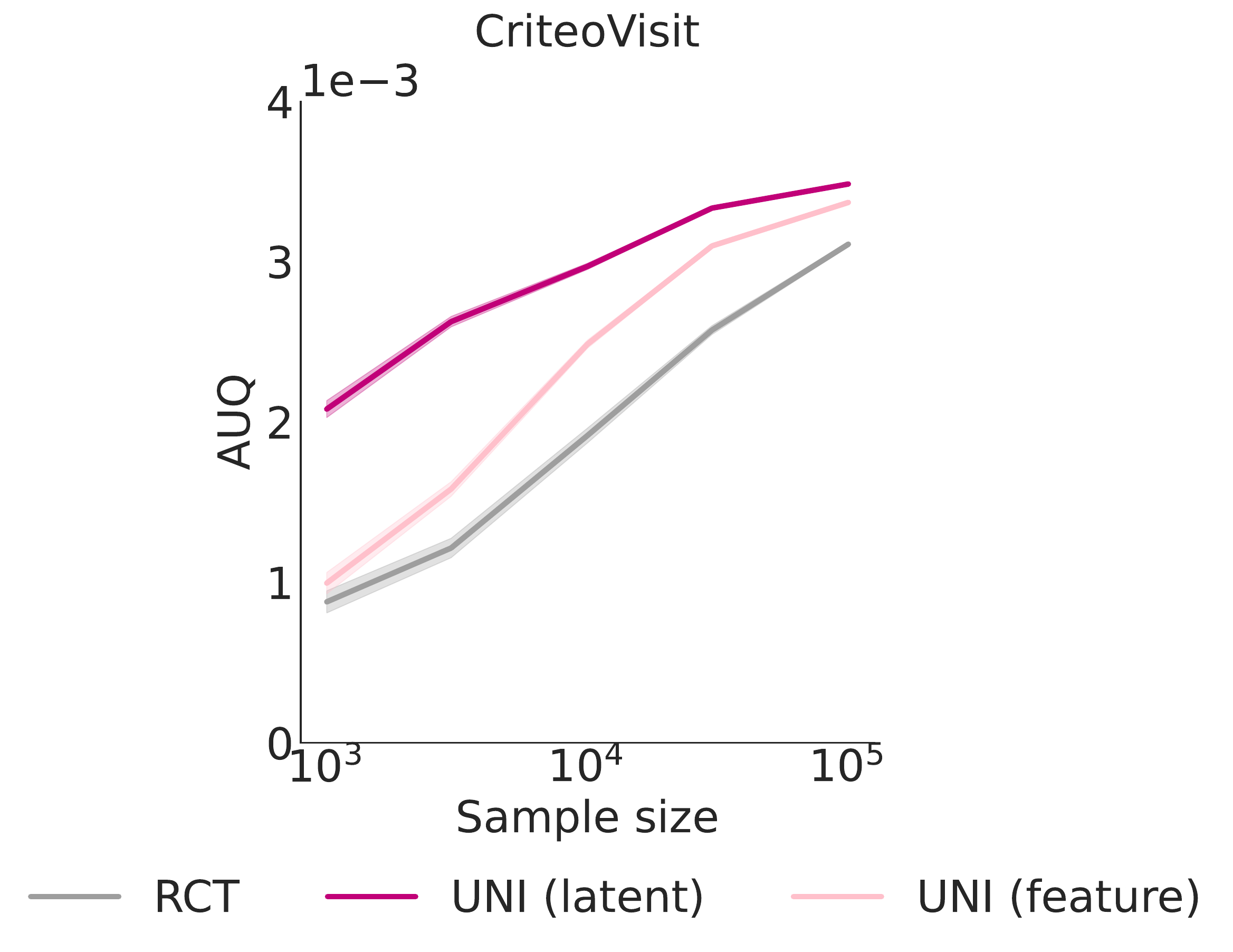}}
    \caption{Comparison of uniform sampling in latent versus feature space on \CriteoVisit.}
\label{fig:feature-space}
\vskip -5mm
\end{center}
\end{figure}

\begin{figure}[t!]
\begin{center}
\centerline{\includegraphics[width=0.8\columnwidth]{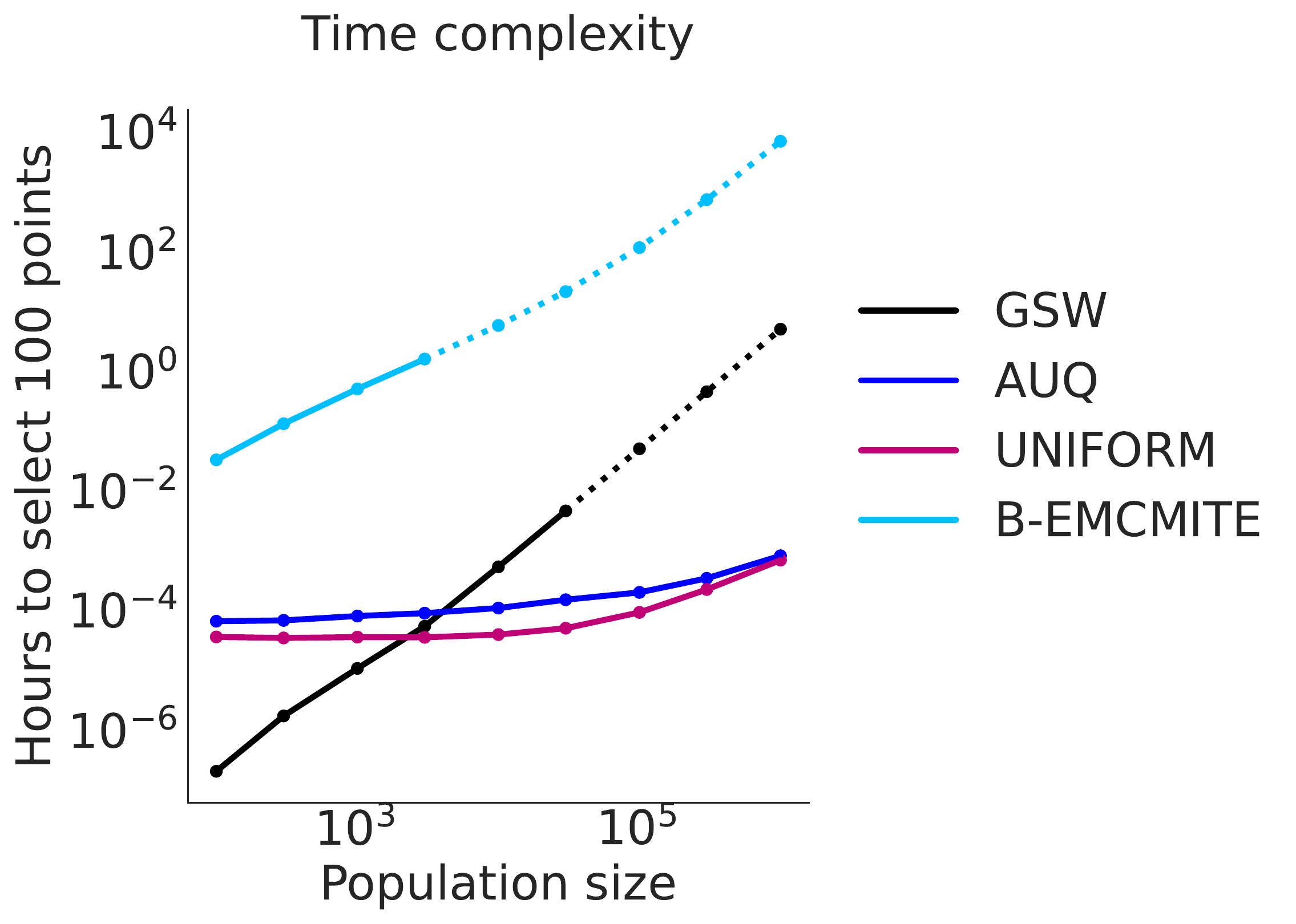}}
\caption{Empirical time complexity of our method benchmarked against Recursive GSW and B-EMCMITE.}
\label{fig:time-complexity}
\vskip -10mm
\end{center}
\end{figure}

\begin{figure*}[t!]
\begin{center}
\centerline{\includegraphics[width=0.8\textwidth]{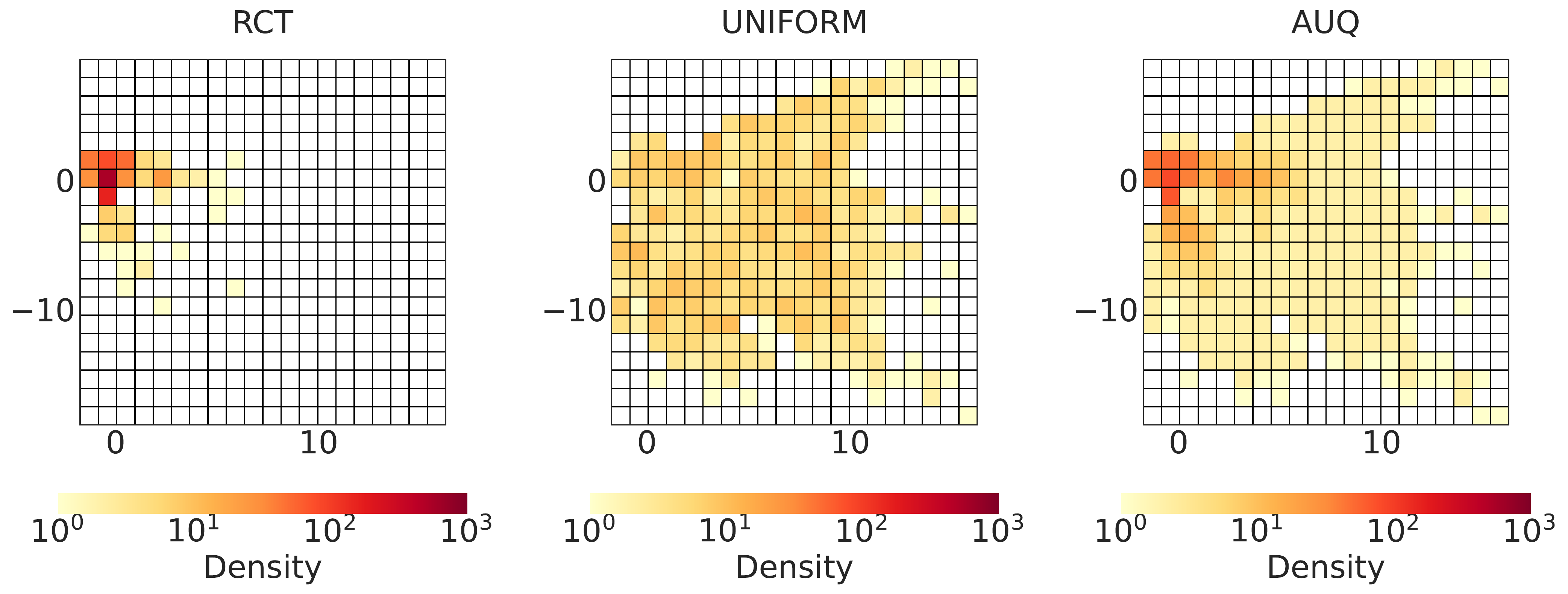}}
\caption{The first $10^3$ points selected from \CriteoVisit by RCT, uniform sampling in latent space, and our AUQ-optimised sampler displayed with respect to the rectangular grid on latent space.}
\label{fig:rct-uni-auq-latentsamples}
\end{center}
\end{figure*}

\begin{figure*}[t!]
\begin{center}
\centerline{\includegraphics[width=0.7\textwidth]{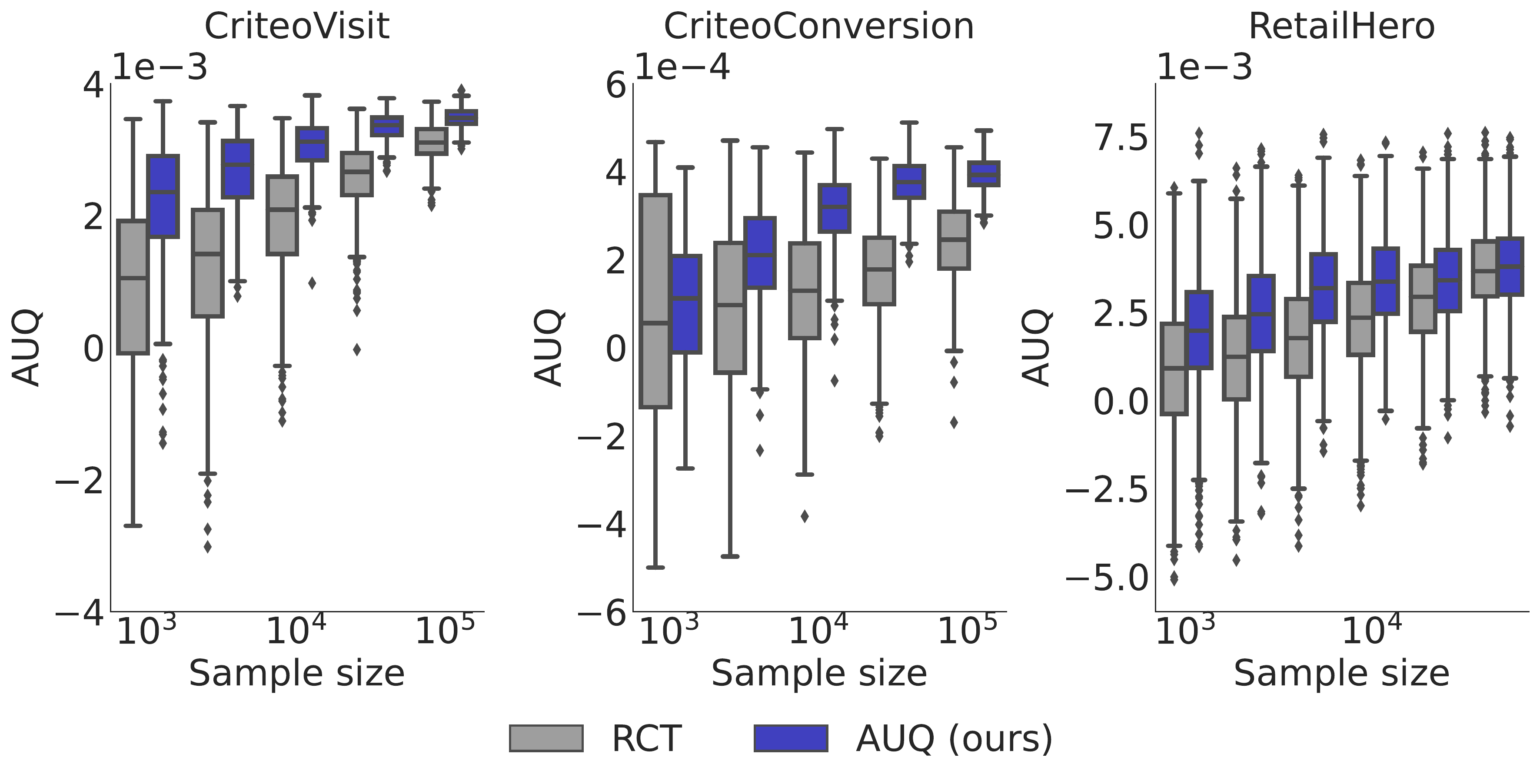}}
\caption{Box-and-whiskers plots showing the distribution in performance achieved by AUQ-optimised sampling and RCT across trials. These are the same experimental results used to compute mean performance in \Fig{marketing}.}
\label{fig:box_whisker}
\end{center}
\end{figure*}

\begin{figure*}[t!]
\begin{center}
\centerline{\includegraphics[width=0.6\textwidth]{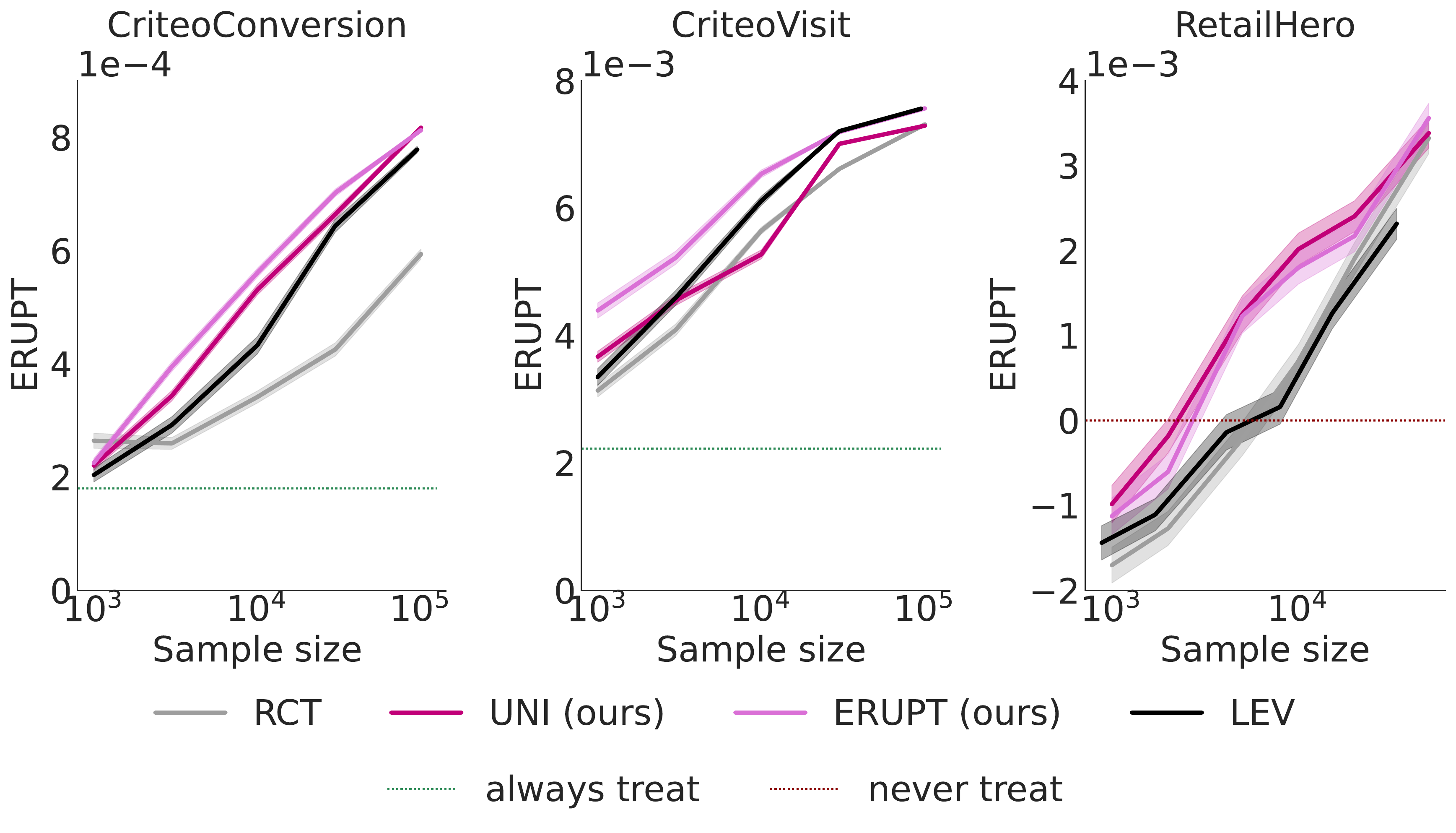}}
\caption{Performance of ERUPT-optimised sampling across datasets, benchmarked against RCT, Leverage Scoring, as well as always-treat and never-treat baselines.}
\label{fig:erupt}
\end{center}
\end{figure*}

\Fig{feature-space} demonstrates the performance of uniform sampling in latent space on \CriteoVisit benchmarked against uniform sampling in \emph{feature space}. For the latter, we split each raw feature at its median value, which results in $2^{n} = 4096$ distinct bins for this dataset with $n=12$ features (though only 238 of them are populated). This baseline performs better than RCT but much worse than uniform sampling in latent space, demonstrating the advantage of the smooth dense representation that the VAE provides.

\Fig{time-complexity} benchmarks the time complexity of our sampling algorithms against Recursive GSW \cite{addanki2022sample} and B-EMCMITE \cite{puha2020batch} as a function of population size. For cases where it was more expensive to run the experiments at the largest population sizes, the data points were extrapolated (dashed line) from the experimentally observed results (solid line). This explains our limited application of these other methods in our experiments and demonstrates an additional advantage of our approach.

\Fig{rct-uni-auq-latentsamples} shows the distributions of data points selected by RCT, uniform sampling in latent space, and our AUQ-optimised sampler. The plots display the occupancies within the rectangular grid on latent space after the first $10^3$ points are drawn by each algorithm. Note the broad similarity between the cohorts selected by AUQ-optimised sampling and uniform sampling in latent space.

\Fig{box_whisker} shows the performance of our AUQ-optimised sampler, benchmarked against RCT, and displayed as a box-and-whiskers plots at each sample size. This gives visibility into the distribution of performance achieved by each method, whereas \Fig{marketing} showed only the mean performance over many trials.

\Fig{erupt} displays the performance of our ERUPT-optimised sampler, which is  presented below in \App{extra_theory}. Our method outperforms Leverage Scoring \cite{addanki2022sample} and RCT across samples sizes and datasets. The ERUPT-optimised sampler also distinctly outperforms uniform sampling in latent space, in contrast to the results of \Fig{marketing} for the AUQ target. We describe both this metric and this experiment in further detail in \App{erupt}.

\begin{figure}[t!]
\begin{center}
\centerline{\includegraphics[width=0.65\columnwidth]{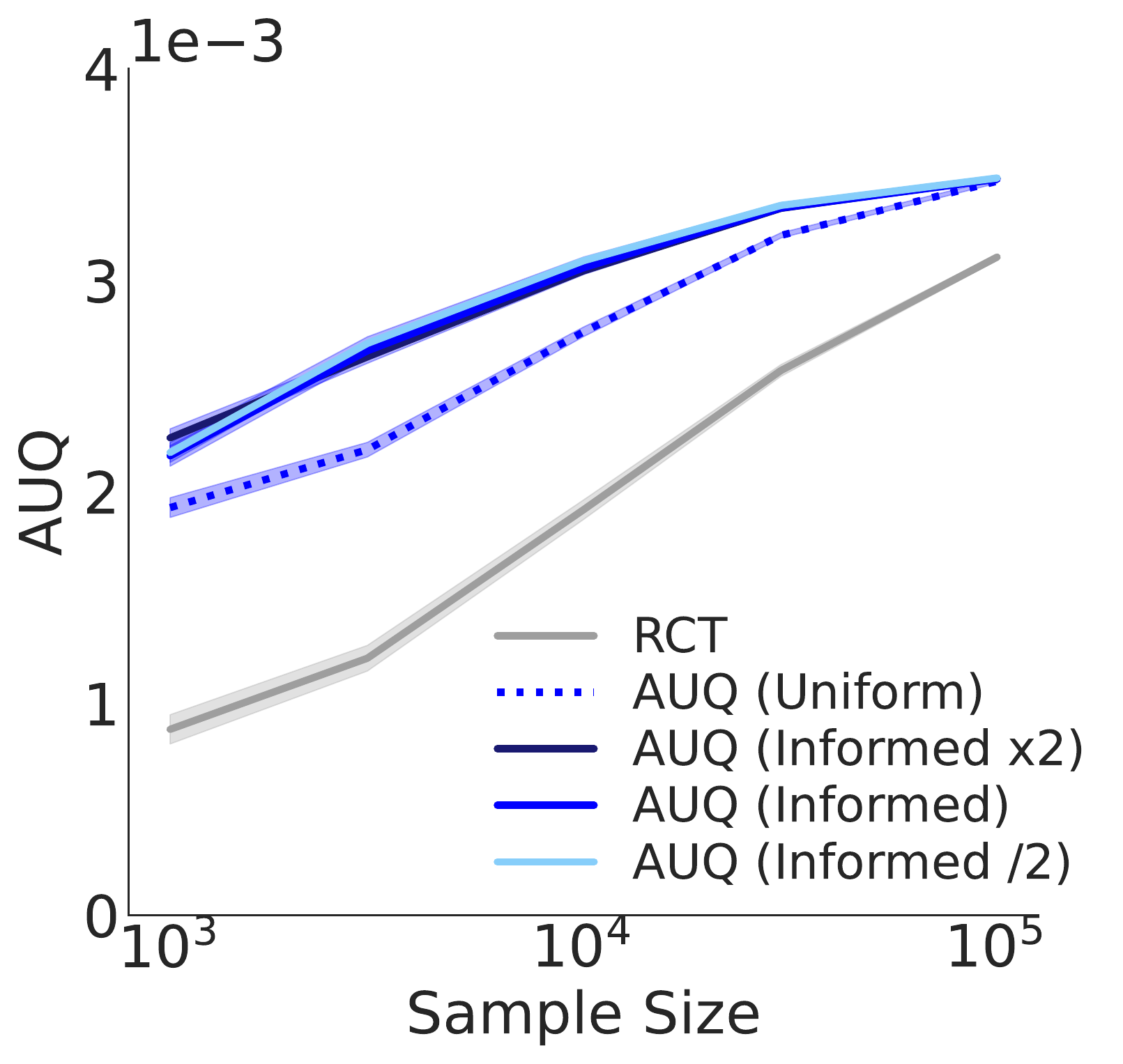}}
\caption{Dependence of AUQ-optimised sampler on the informed prior's hyperparameters on \CriteoVisit.}
\label{fig:informed_prior_sensitivity}
\vskip -5mm
\end{center}
\end{figure}

\Fig{informed_prior_sensitivity} explores the sensitivity of the informed prior discussed in \Sec{hyperparams} to inaccurate estimates of the treatment effect by systematically over- and under-estimating $n_t^\pm$ while keeping $n_t^+ + n_t^-$ constant. The $n_t^\pm$ values can be combined to predict the ATE, and we varied them in such a way as to predict both double and half the actual ATE in the test set. \Fig{informed_prior_sensitivity} demonstrates that our optimised samplers are robust to such substantial variations in the prior.

\begin{figure}[t!]
\begin{center}
\centerline{\includegraphics[width=\columnwidth]{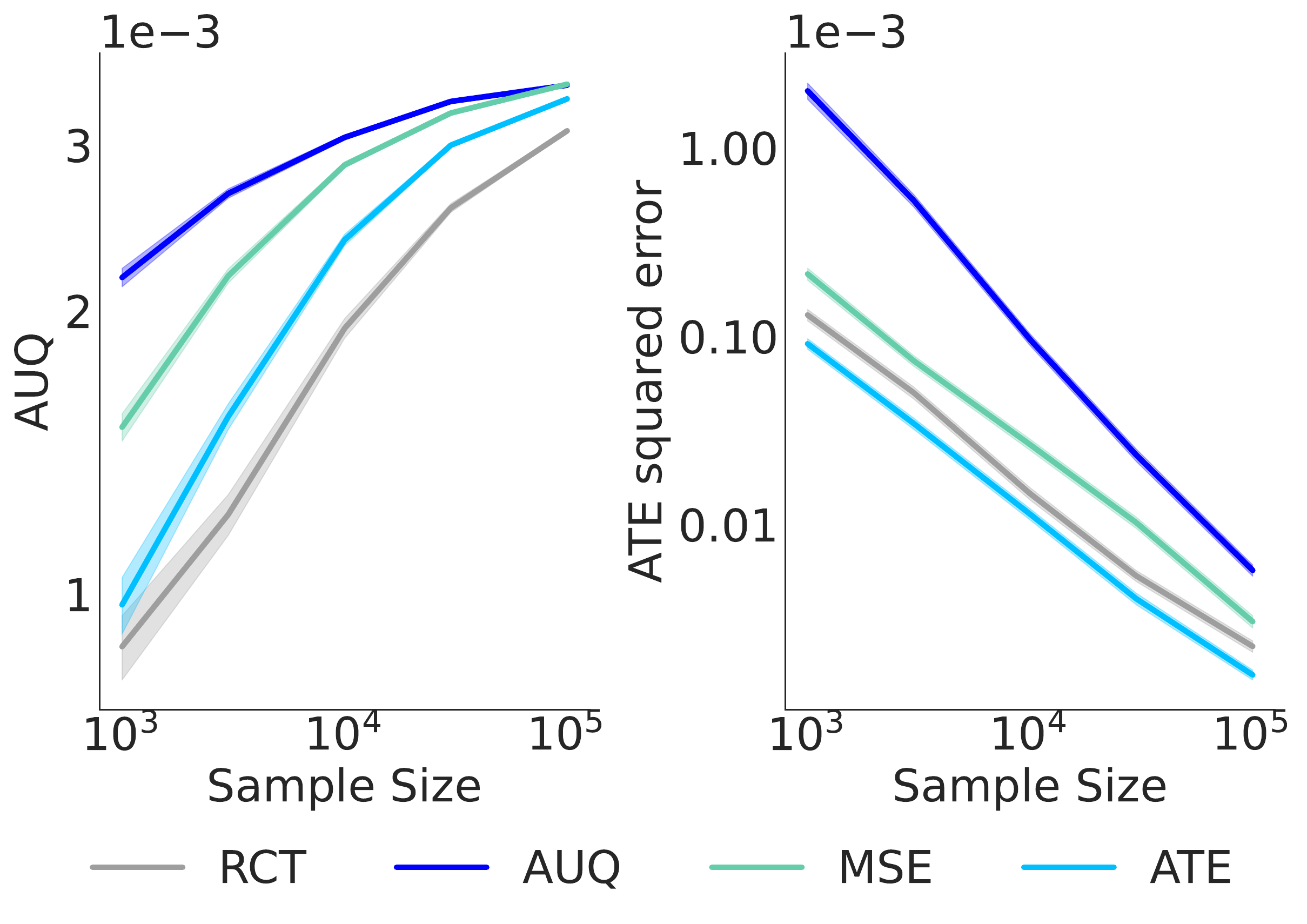}}
\caption{Performance of AUQ\,/\,ATE\,/\,MSE\,-\,optimised samplers across target metrics on \CriteoVisit.}
\label{fig:task-specificity}
\vskip -5mm
\end{center}
\end{figure}

\Fig{task-specificity} displays the performance of our AUQ-, ATE-, and MSE-optimised samplers evaluated on both the AUQ and ATE-squared-error metrics on \CriteoVisit. In a sense, this result shows that one cannot ignore the downstream task for which an experiment is being designed. No sampler, not even RCT, performs well across target metrics as disparate as AUQ and ATE squared error.

\begin{figure*}[ht]
\begin{center}
\centerline{\includegraphics[width=0.8\textwidth]{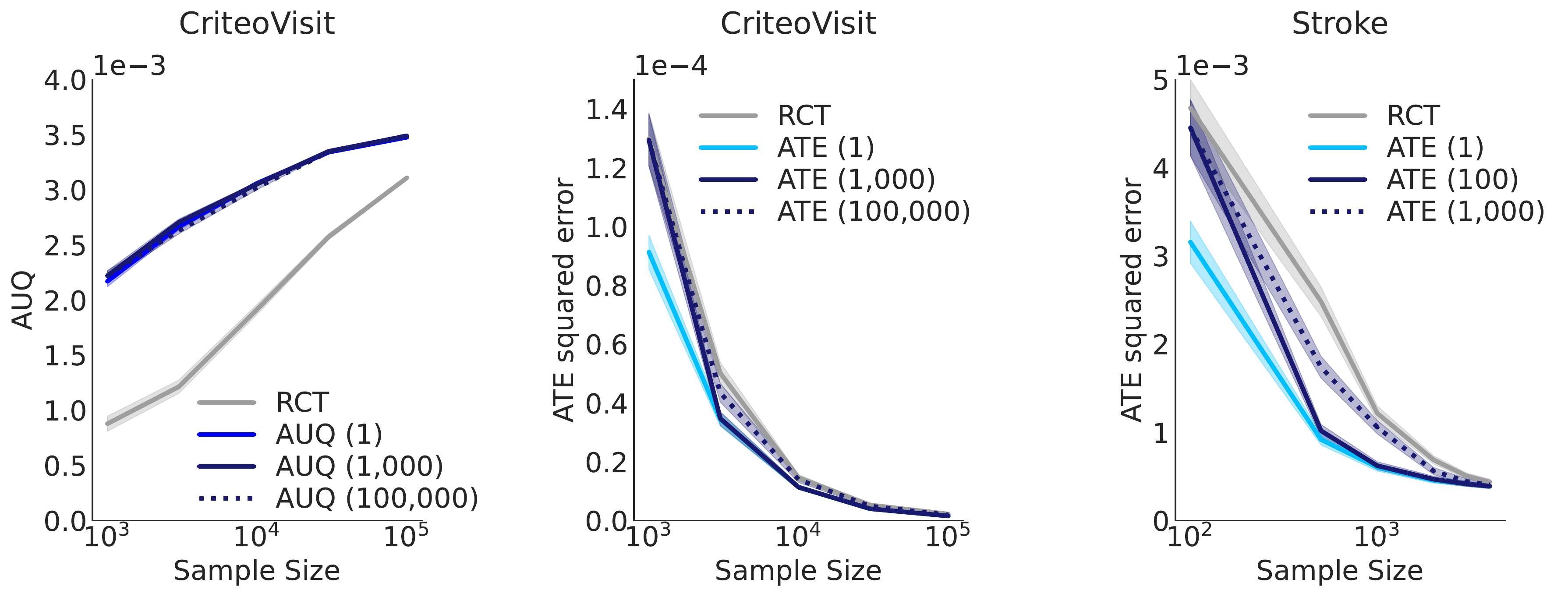}}
\caption{Performance of optimised samplers across datasets and target metrics when our method is run in batch mode (with batch sizes listed in parentheses in the legends).}
\label{fig:batch}
\vskip -5mm
\end{center}
\end{figure*}

\Fig{batch} displays the performance of our optimised samplers across datasets when our method is run in \emph{batch mode}. Referring to the 4-step procedure of \Sec{our_method}, batch mode consists of iterating between steps 1-2 (until a full batch is selected) and deferring steps 3-4 to be performed at longer intervals (between batches). This reduces the frequency at which the posterior distribution for Thompson sampling is updated with new observations. Batch sizes are indicated in parentheses in the legends of \Fig{batch}. This result shows positive performance even at large batch sizes.

\section{Experimental details}
\label{app:experimental}

Here we provide the details of our setup for each of our empirical analyses.

\subsection{Datasets}
\label{app:datasets}

\subsubsection*{Synthetic Data}

We created 1-dimensional synthetic datasets for \Sec{intuition} by sampling continuous features from a truncated normal distribution on $[0,1]$, with mean $\mu=0$ and $\sigma=0.2$:
\eqn{
p(X=x) = \frac{1}{\sigma} \cdot \frac{\varphi\left(\frac{x-\mu}{\sigma}\right)}{\Phi\left(\frac{1-\mu}{\sigma}\right) -\Phi\left(\frac{-\mu}{\sigma}\right)  }
}
where $\varphi$ denotes the probability density function of the standard normal distribution, and $\Phi$ is its corresponding cumulative distribution function. The data density is plotted in \Fig{intuition}(a). We set the baseline propensity to be a linear function of the features 
\eqn{
    p(Y=1 | X=x, T=0) = 0.4 \, x
}
The baseline propensity is plotted in \Fig{intuition}(b). We explored two sigmoidal uplift distributions, designed to replicate the situation where the largest uplift individuals are found in either the tail or the bulk of the data:
\eqn{
    u_i(x) = \frac{a_i}{ (1 + \exp(-b_i (x - c_i)))}
}
where $(a_i,b_i,c_i) = (0.7, \pm 20, 0.2)$. The two uplift functions are plotted in \Fig{intuition}(c). Target metrics in \Fig{synthetic_performance} were computed on a test set of 4M points.
 
\subsubsection*{Stroke}

We extracted this dataset from the International Stroke Trial \cite{sandercock2011international}, a large scale randomised trial studying the health outcomes of 19,435 stroke patients given different treatments upon hospital admission. We extracted 40 patient level features such as age, sex, and blood pressure which were recorded upon admittance to the hospital. To study a binary treatment effect we used patients receiving no treatment drug as the control group and patients receiving only aspirin as the treatment group. We also dropped rows from the pilot phase of the trial, as well as patients who received alternative treatments, to get a dataset of 9,208 rows. The binary outcome corresponds to whether or not the patient was discharged alive from hospital.

\subsubsection*{CriteoVisit \& CriteoConversion}

The Criteo Uplift Modelling dataset \cite{Diemert2018} is a large scale benchmark for ITE estimation. It has 13,979,592 rows comprised of 12 anonymized customer features, a binary treatment indicator, and two possible binary outcome measures (Visit and Conversion). We used this dataset twice (\CriteoVisit and \CriteoConversion) to consider each outcome in turn. The public treatment group was randomly down-sampled to give a 50/50 balance between treatment and control groups. This resulted in a train pool of 2,992,640 rows and a test set of 4,000,000 rows.

\subsubsection*{RetailHero}

We created this dataset from raw customer, sales, and marketing trial data in the scikit-uplift X5 RetailHero dataset\footnote{Raw data: \url{https://www.uplift-modeling.com/en/v0.3.1/api/datasets/fetch_x5.html}} which contains raw information about previous purchases made by customers of the X5 RetailGroup. The customers were exposed to a binary treatment and their corresponding binary outcome was recorded. Customer features (e.g.~age and gender) were combined with new features engineered from their purchase history e.g.~total historical spend, average per transaction spend, and number of stores visited. This resulted in a dataset of 23 customer features with 100,036 train rows and 100,000 test rows. We have made this dataset and information about the engineered features publicly available.\footnote{Processed data: \url{https://tinyurl.com/RetailHero}}

\subsection{Models}

\subsubsection*{VAE}

The VAE architecture we used in our experiments is comprised of a 2-layer fully-connected encoder with 100-dimensional hidden layers, a 2-dimensional latent space, and a decoder with the same architecture as the encoder which outputs a Bernoulli probability for binary features, a softmax for categorical features, and a mean \& variance for continuous features. Given the positive results this simple architecture supported, we did not tune the VAE beyond this vanilla baseline \cite{kingma2013auto}. We trained the VAE using Adam \cite{kingma2014adam} with learning rate $10^{-4}$ and early stopping on the validation-set ELBO.

\begin{figure*}[ht]
\begin{center}
\centerline{\includegraphics[width=\textwidth]{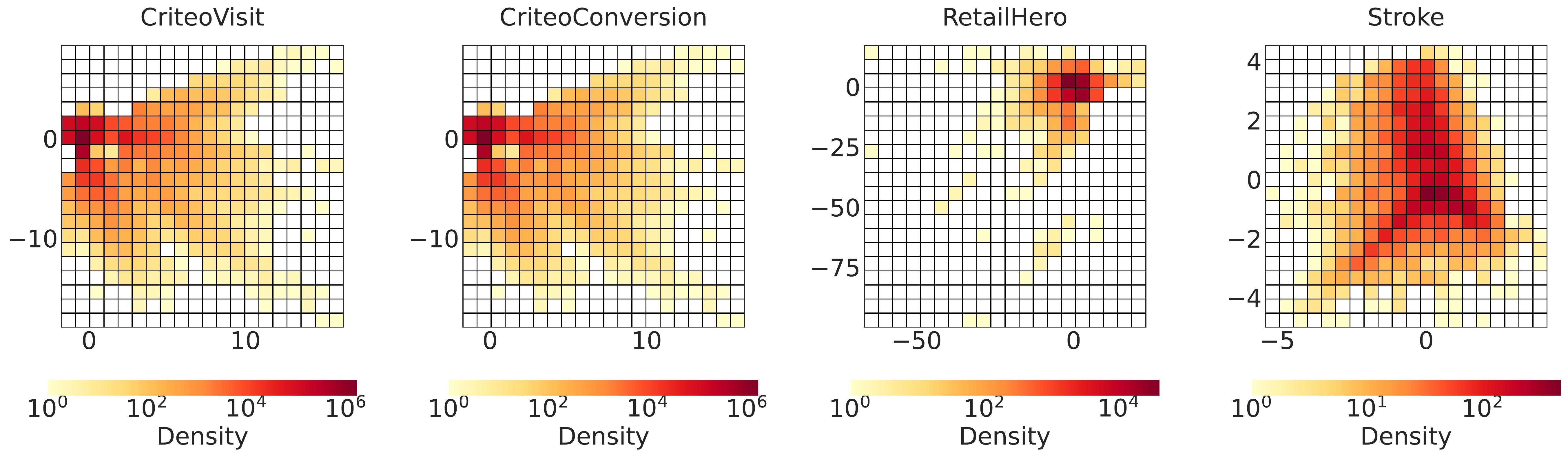}}
\caption{Discretised latent representations of the full population of each dataset studied in this paper.}
\label{fig:Datasets Latent Density and Uplift Distributions}
\end{center}
\end{figure*}

The first step of our algorithm (see \Fig{schematic}) is to discretise the population of training data in latent space. We discretised the continuous latent representation of each dataset by taking the smallest hypercube that contains all of the data and slicing each edge of the hypercube uniformly into a number of cells (with a default of 20 unless stated otherwise). \Fig{Datasets Latent Density and Uplift Distributions} provides a visualisation of the resulting latent distribution for each dataset studied in this paper.

\subsubsection*{ATE}

For treatment effect estimation when targeting the ATE, we began by computing the empirical uplift in each grid cell $b$ of discretised latent space $\mathcal B$ according to the sampled training data. We then estimated the ATE by weighting each of these empirical uplift values by the data density in that grid cell and summing:
\eqn{
    \widehat{\text{ATE}} &= \sum_{b \in \mathcal{B}} \, p(b) \, \sum_{t = 0}^1 \, \sum_{i=1}^{n(b,t)}  \frac{(-1)^{t+1}}{n(b, t)} \, Y_{i(b, t)}
}
Here the density $p(b)$ in each cell is calculated according to our available pool of training samples.

\subsubsection*{ITE}

For ITE estimation, we used custom implementations of the T-learner and the S-learner \cite{kunzel2019metalearners}. We defaulted to the T-learner in all our experiments unless stated otherwise (e.g.~in \Fig{models}). For either model, the core learners of the ITE estimator were XGBoost models initialised with following hyperparameters: 
\begin{itemize}
    \item \texttt{n-estimators}: $400$\\[-20pt]
    \item \texttt{objective}: \texttt{binary:logistic}\\[-20pt]
    \item \texttt{eval-metric}: \texttt{rmse}\\[-20pt]
    \item \texttt{max-depth}: 1 (T-learner), ~2 (S-learner)
\end{itemize}
 Additionally during model training, the sampled data was partitioned 80/20 into training/validation sets for early stopping (with \texttt{early-stopping-rounds}: $50$).

\subsection{Benchmarks}

We benchmarked our optimised samplers against the work of \citet{addanki2022sample} which provides two sample efficient algorithms for ATE and ITE estimation, namely Recursive GSW and Leverage Scoring, respectively. The code for using these algorithms on synthetic or semi-synthetic data, where both treatment outcomes are available, is open source\footnote{\url{https://github.com/raddanki/Sample-Constrained-Treatment-Effect-Estimation}}. We modified this code in order to benchmark the methods on real data. Besides the modifications below, the algorithms were implemented exactly as reported. 

The Recursive GSW algorithm assigns treatments by recursively splitting a given pool of training data $\mathcal{D}_{\text{pool}}$ into halves 
($\mathcal{D}_{\text{treatment}}$ and $\mathcal{D}_{\text{control}}$) with covariates balanced between the two. The aim is to obtain a small treatment and control group which are both balanced and representative of $\mathcal{D}_{\text{pool}}$ as a whole. Since our work utilises large datasets originally obtained by RCT, we treated the two available treatment arms as independent pools of data which are already balanced. We then performed the Recursive GSW algorithm on each treatment group separately to obtain treatment and control samples.

We did not run Recursive GSW for \CriteoVisit and \CriteoConversion because the algorithm scales poorly to large datasets. See \Fig{time-complexity} for a visualisation of time for one split of a dataset of size $n_\text{pool}$.

The Leverage Scoring algorithm works by assigning sampling probabilities $\pi$ to each point in $\mathcal{D}_{\text{pool}}$. The treatment and control groups are then independently sampled with probability $\pi$ for each data point. 
To replicate this on our datasets, we again treated the available treatment and control groups as two independent pools of data with independent $\pi$. Treatment and control groups were then independently sampled as above.

Due to the stochastic nature of these batch sampling algorithms, the desired sample size is not guaranteed. Instead of reporting performance of these samplers at the target sample size (as in the original paper), we report it at the average sample size gathered over many experimental seeds which, in the case of Leverage Scoring,  we found to be systematically smaller than the target size. Furthermore, for Recursive GSW we gathered sample sizes only of particular fractions of the population size ($n_\text{pool}/2^s$). While it is possible to collect intermediate sample sizes by random sub-sampling, this would introduce randomisation and degrade the covariate balancing that the algorithm is designed to achieve.

\subsection{Compute and empirical analysis}

All experiments were performed in parallel on 96 core, 393 GB machines. For all datasets except \Stroke, we performed 384 trials per experiment, and we bootstrap-resampled the test set for each trial. Because of its smaller size, experiments on \Stroke each consisted of 1000 trials, and we performed a fresh train-test split for each trial. We reported the performance of each sampling method as the mean target metric achieved across these trials, with uncertainty bands representing the standard error.

\section{Theoretical details}
\label{app:extra_theory}

In this appendix, we provide the algorithms, derivations, and additional details mentioned throughout the paper. We begin by providing the algorithm for evaluating the Qini curve on an RCT test set in \Alg{qini}.

\begin{algorithm}[t!]
\caption{Qini curve estimation on RCT test set}\label{alg:qini}
\begin{algorithmic}
\REQUIRE 
Data $\mathcal{D} = \{(x_i, t_i, y_i)\}_{i=1}^n$ \\[5pt]
Uplift model $\hat{u}$ \\
Number of points $m$ at which to approximate curve \\[5pt]
\ENSURE 
Qini curve values $\{(f_i, q_i)\}_{i=0}^{m}$ \\[5pt]
\STATE Sort $\hat{\mathcal{D}} = \{(\hat{x}_i, \hat{t}_i, \hat{y}_i)\}_{i=1}^n$ so $\hat u(\hat x_i) \geq \hat u(\hat x_{i+1})$ for all $i$ \\
\STATE Set $(f_0, q_0) = (0, 0)$ \\[5pt]
\FOR{$i=1$ \TO $m$}
\STATE $k = \lfloor n \cdot i / m \rfloor$ \\[5pt]
\STATE $k_0 = \sum_{j=1}^k (1 - \hat t_j)$ \\[5pt]
\STATE $k_1 = \sum_{j=1}^k \hat{t}_j$ \\[5pt]
\STATE $u = k_1^{-1} \sum_{j=1}^k \hat{t}_j \, \hat{y}_j ~-~ k_0^{-1} \sum_{j=1}^k (1 - \hat{t}_j) \, \hat{y}_j $ \\[5pt]
\STATE Set $(f_i, q_i) = (i / m, \, u \cdot i / m)$ \\[5pt]
\ENDFOR
\RETURN $\{(f_i, q_i)\}_{i=0}^{m}$
\end{algorithmic}
\end{algorithm}

\subsection{The ERUPT metric}
\label{app:erupt}

ERUPT\footnote{Coined by S.~Weiss (2019) in an eponymous \href{https://medium.com/building-ibotta/erupt-expected-response-under-proposed-treatments-ff7dd45c84b4}{Medium article}.} (Expected Response Under Proposed Treatments) \cite{zhao2017uplift, hitsch2018heterogeneous} quantifies a model's performance at the downstream task of predicting the optimal treatment to assign each member of the population. It is defined as follows.

Given an uplift model $\hat{u}: \mathcal{X} \to \mathcal{Y}$ for a binary treatment, we can use this model to decide whether to treat a member of the population $x$ according to the rule:
\begin{equation*}
\text{treat} ~~x~~ \text{if} ~~ \hat{u}(x) > c
\end{equation*}
where $c$ denotes the threshold for treatment being ``worthwhile.'' For example, if the outcome $\mathcal{Y}$ is continuous and reflects spend, this can be interpreted as the cost of the treatment (over the baseline or control). The ERUPT metric measures the performance of such assignments by calculating the true uplift $u$ (minus cost $c$) over those who were assigned treatment by the model:
\begin{equation}\label{eqn:erupt}
    \text{ERUPT}[\hat{u}] = \int_{\mathcal{X}} p(x) \, \Theta(\hat{u}(x) - c) \, (u(x) -c) \, dx
\end{equation}
where $\Theta$ is the step function defined below in \Eq{theta-fn}. Larger values of ERUPT correspond to better treatment assignments by the model. ERUPT is maximal when $\hat u = u$.

In the experimental results of \Fig{erupt}, the ``cost'' of treatment $c$ is taken to be the ATE. This choice is motivated by the fact that with such a cost, a random baseline would break even at ERUPT$=0$. \Fig{erupt} also marks the performance of ``always treat'' and ``never treat'' baselines as well.

\subsection{Bias and variance of \Eq{bayesian_model}}

In the following sections, we provide derivations of the core functions $f_\alpha^{\mathcal R}\big(\{n(x, t)\} \,;\, \theta^* \big)$ that appear in \Sec{our_method} and around which our method is built. We will perform these calculations explicitly for the case in which the population is partitioned into bins to ensure that the discretisation process within our method does not introduce any bias into our approach. So in this section, we begin with the derivation of the bias and variance results of \Eq{bias_var}, but under the assumption that the feature space $\mathcal{X}$ has been discretised into $k$ bins. Let $\mathcal{B}$ represent the set of these bins. The model defined in \Eq{bayesian_model} applied to bins $b\in \mathcal{B}$ then becomes:
\eqn{
\hat U(b) ~&=~ \frac{\alpha_{b1} ~+~ \sum_{i=1}^{n(b, 1)} \,Y_{i(b, 1)}}{\alpha_{b1} ~+~ \beta_{b1} ~+~ n(b, 1)} \nonumber \\[5pt] 
&\qquad~-~ \frac{\alpha_{b0} ~+~ \sum_{i=1}^{n(b, 0)} \,Y_{i(b, 0)} }{\alpha_{b0} ~+~ \beta_{b0} ~+~ n(b, 0)} \label{eq:ball_model}
}
where each $Y_{i(b, t)} \sim p(y | X=X_{i(b,t)}, T=t)$ and each $X_{i(b,t)} \sim p|_b$. Instead of estimating the uplift at a fixed $x\in \mathcal{X}$, this is now an estimate of the \emph{average} uplift in $b$:
\eqn{\nonumber
u(b) &= \frac{1}{p(b)} \int_{b}  \big(\e{Y| X=x, t=1} \\
& \hspace{15mm} -~  \e{Y| X=x, t=0}\big) \, p(x)  \, dx \label{eq:avuplift}
}
where
\eqn{
    p(b) & := \int_{b} p(x)\, dx
    }

To see this, we calculate the expected value of $\hat{U}(b)$ in the bin $b$. Since $\hat{U}(b)$ depends on random variables in both $X$ and $Y$, we apply the law of total expectation to see that
\eqn{\nonumber
        \e{\sum_{i=1}^{n(l,t)} Y_i|_{X=X_i, t}} &=\frac{n(b,t)}{p(b)}\int_{b} \e{Y| X=x, t} p(x) dx \\ \label{eq:expected}
        &=: n(l,t)\e{Y|b, t}
    }
and so 
\eqn{
        \e{\hat{U}(b)} &= \frac{\alpha_{b1} ~+~ n(b,1) \, \e{Y|b, t=1}}{\alpha_{b1} ~+~ \beta_{b1} ~+~ n(b,1)} \nonumber \\[5pt]
        & - \frac{\alpha_{b0} ~+~ n(b,0) \, \e{Y|b, t=0}}{\alpha_{b0} ~+~ \beta_{b0} ~+~ n(b,0)} 
    }
We see from this expression that in the limit of large $n(b,t)$ we have $\mathbb{E}[\hat{U}(b)] \to u(b)$. Similarly, we can apply the law of total variance to calculate
\eqn{
        \var{\sum_{i=1}^{n(b,t)} Y_i|_{X=X_i, t}} &= \frac{n(b, t)}{p(b)}\int_{b} \var{Y| X=x} p(x) dx \nonumber \\ 
        &+ n(b,t)\, \var{\e{Y|X}}
    }
where we have leveraged the pairwise independence of samples $(X,Y)$. It transpires that
\eqn{
        \var{\sum_{i=1}^{n(b,t)} Y_i|_{X=X_i, t}} = n(b, t) \, \sigma^2(b,t)
    }
where
\eqn{\label{eq:sigma_cts}
    \sigma^2(b,t) &:= \e{Y^2| b, t} - \e{Y|b, t}^2
    }
So we can write
\eqn{
        \var{\hat{U}(b)} &= \frac{n(b, 1) \sigma^2(b,1)}{(\alpha_{b1} ~+~ \beta_{b1} ~+~ n(b,1))^2} \nonumber \\[5pt]
        &+ \frac{n(b, 0) \sigma^2(b,0)}{(\alpha_{b0} ~+~ \beta_{b0} ~+~ n(b,0))^2}
    }

Writing $\theta^*_{bt} = \e{Y|b, t}$, we see that when $\mathcal{Y}=\{0,1\}$ we have
\eqn{
&\mathbb E[\hat U(b) - u(b)] ~=~ \frac{\alpha_{b1} - (\alpha_{b1} + \beta_{b1}) \, \theta^*_{b1}}{\alpha_{b1} + \beta_{b1} + n(b, 1)} ~-~ (1 \to 0) \nonumber \\[5pt]
&\mathbb V[\hat U(b)] ~=~ \frac{\theta^*_{b1}\,(1-\theta^*_{b1})\,n(b, 1)}{\big(\alpha_{b1} + \beta_{b1} + n(b, 1)\big)^2} ~+~ (1 \to 0) \label{eq:cts_bias_var}
}
in agreement with \Eq{bias_var}, though with a slightly altered interpretation of $\theta^*$. The model of \Eq{ball_model} then induces the following model on $\mathcal{X}$:
\begin{equation}\label{eq:cts_bayesian_model}
    \hat{U}(x) = \sum_{b\in \mathcal{B}} \hat{U}(b)\chi(x\in b)
\end{equation}
where $\chi$ denotes the indicator function.

\subsection{MSE-optimised sampler}

Next we derive \Eq{f_mse}, but for the case in which we have discretised $\mathcal X$ as described above. In this case, the MSE-optimised sampler in fact aims to minimise the density-weighted squared error of the predicted average uplift in each bin $b \in \mathcal B$, rather than the squared error in uplift at each point $x \in \mathcal{X}$. In particular,
\eqn{
\mathbb E\big[ \text{MSE}[\hat U] \big] &= \sum_{l=1}^k \, p(b) ~ \mathbb E\big[(\hat{U}(b) - u(b))^2\big] \nonumber \\[5pt]
&= \sum_{b\in\mathcal{B}} \, p(b) \, \Big(\mathbb E\big[\hat U(b) - u(b)\big]^2 + \mathbb V\big[\hat U(b)\big] \Big) \nonumber \\[6pt]
&=: -f_\alpha^\text{MSE}\big(\{n(b, t)\};\,\theta^*\big)
}
where dependence on prior parameters, $n(b,t)$, and $\theta^*$ can be inferred from \Eq{cts_bias_var}. 

\subsection{ATE-optimised sampler}
 
Next we derive the discretised analog of \Eq{f_ate} for the ATE-squared error:
\eqn{
    &\mathbb E\Big[\big(\text{ATE}[\hat{U}] - \text{ATE}[u]\big)^2\Big] \nonumber \\[5pt]
    &= \mathbb E\big[ \text{ATE}[\hat U] - \text{ATE}[u] \big]^2 + \mathbb V\big[\text{ATE}[\hat U]\big] \nonumber \\[8pt]
    &= \mathbb E\Big[ \int_{\mathcal X} \big(\hat U(x) - u(x)\big) p(x) \,dx \Big]^2 \nonumber \\[5pt]
    &+ \mathbb V\Big[\int_{ \mathcal X} \hat U(x)p(x)\, dx \Big] \nonumber \\[5pt]
    &= \mathbb E\Big[ \sum_{b \in \mathcal{B}}\big( p(b) \hat U(b) - \int_{ b} u(x) p(x) \,dx \big) \Big]^2 \nonumber \\[5pt]
    &+ \mathbb V\Big[\sum_{b \in \mathcal{B}} \hat U(b)p(b) \Big] \nonumber \\[5pt]
    &= \Big(\sum_{b\in\mathcal{B}}\, p(b) \, \mathbb E\big[ \hat U(b) - u(b) \big] \Big)^2 + \sum_{b\in\mathcal{B}} \, p(b)^2 \, \mathbb V\big[\hat U(b)\big] \nonumber\\[5pt]
    &=: -f_\alpha^\text{ATE}\big(\{n(b, t)\};\,\theta^*\big)
}
where dependence on prior parameters, $n(b,t)$, and $\theta^*$ can be inferred from \Eq{cts_bias_var}.

\subsection{AUQ-optimised sampler}

Next we derive \Eq{f_auq} for the AUQ in the large $n(b, t)$ limit. We first introduce some notation. Let
        \begin{align} \label{eq:theta-fn}
            \Theta(s) = \left\{ \begin{array}{ll}
                0 & \text{if $s < 0$} \\
                1/2 & \text{if $s = 0$} \\
                1 & \text{if $s > 0$}
            \end{array}\right.
        \end{align}
        Then we can write
        \begin{align}
            \textnormal{AUQ}[\hat U] &= \sum_{\substack{b, b'\in \mathcal{B}\\ b'\ne b}}  p(b) \, p(b') \, u(b') \, \Theta(\hat{U}(b') - \hat{U}(b)) \nonumber \\[5pt]
            &-\frac{1}{2}\sum_{b \in \mathcal{B}} p(b) \, u(b)
        \end{align}
which implies
    \begin{align}
        \mathbb E\big[\textnormal{AUQ}[\hat U]\big] &= \sum_{\substack{b, b'\in \mathcal{B}\\ b'\ne b}} p(b)\, p(b')\, u(b') \, \e{\Theta(\hat{U}(b') - \hat{U}(b))} \nonumber \\[5pt]
        &-\frac{1}{2}\sum_{b\in \mathcal{B}} p(b) \, u(b)
    \end{align}
We thus need to compute
    \begin{equation}
         \e{\Theta(\hat{U}(b') - \hat{U}(b)) }
    \end{equation}
    We will not compute this exactly, but the central limit theorem tells us that for large $n(b,t)$ and $n(b',t)$, the difference $\hat{U}(b') - \hat{U}(b)$ is approximately normally distributed, with mean and variance given by
    \begin{align}
        \mu(b', b) &= u(b') - u(b) \\[5pt]
        \sigma^2(b', b) &= \frac{\sigma^2(b', 1)}{n(b', 1)} + \frac{\sigma^2(b', 0)}{n(b',0)} + \frac{\sigma^2(b, 1)}{n(b, 1)} + \frac{\sigma^2(b, 0)}{n(b,0)} \nonumber
    \end{align}
    where we have used the notation \Eq{sigma_cts}. (The dependence on the prior parameters $\alpha_{bt}$ and $\beta_{bt}$ drops out from $\hat U(b)$ in the large $n(b,t)$ limit.) Under this assumption,
    \begin{align}
        \e{\Theta(\hat{U}(b') - \hat{U}(b))} &\approx \int_{-\infty}^\infty \Theta(z) \, \varphi_{\mu, \sigma}(z) \, dz  \\[5pt]
        &= \int_0^\infty \varphi_{\mu, \sigma}(z) \, dz \nonumber \\[5pt]
        &= \frac{1}{2} + \frac{1}{2}\textnormal{erf}\left(\frac{\mu(b', b)}{\sqrt{2\sigma^2(b', b)}}\right) \nonumber 
    \end{align}
    where erf denotes the error function , and we use $\varphi_{\mu, \sigma}$ to denote the normal density function with mean $\mu(b', b)$ and variance $\sigma^2(b',b)$. It follows that
    \eqn{
        &\mathbb E\big[\textnormal{AUQ}[\hat U]\big] \approx \nonumber \\[5pt]
        &\sum_{\substack{b, b'\in \mathcal{B}\\ b'\ne b}} p(b)\, p(b') \, u(b')\left[\frac{1}{2} + \frac{1}{2}\,\textnormal{erf}\left(\frac{u(b') - u(b)}{\sqrt{2 \, \sigma^2(b', b)}}\right)\right] \nonumber \\[5pt]
        &-\frac{1}{2}\sum_{b \in \mathcal{B}} p(b)\, u(b). \nonumber
        }
    in agreement with \Eq{f_auq} up to a constant independent of $n(b, t)$ that can be dropped.

\subsection{ERUPT-optimised sampler}
\label{app:erupt_deriv}

Finally we derive $f_\alpha^{\mathcal R}\big(\{n(x, t)\} \,;\, \theta^* \big)$ for the ERUPT metric, which turns out to be similar to the case of AUQ above:
\begin{align}
    \mathbb{E}[\text{ERUPT}[\hat{U}]] &= \sum_{b \in\mathcal{B}} \, \mathbb{E}\left[\Theta(\hat{U}(b) - c)\right] \int_b (u(x) -c) \,p(x) \,dx
 \nonumber \\[5pt]
 &= \sum_{b \in\mathcal{B}} \, p(b)\, \mathbb{E}\left[\Theta(\hat{U}(b) - c)\right] (u(b) -c) \nonumber 
\end{align}
In the large $n(b,t)$ limit, the central limit theorem implies that $\hat{U}(b) - c$ is approximately normally distributed, with mean $\mu_b = u(b) - c$ and variance
\begin{align}
    \sigma^2_b &= \frac{\sigma^2(b, 1)}{n(b,1)} + \frac{\sigma^2(b, 0)}{n(b,0)}
\end{align}
Thus, for each $b\in \mathcal{B}$, we have that
\begin{align}
   \mathbb{E}\left[\Theta(\hat{U}(b) - c)\right] &= \int_{-\infty}^\infty  \Theta(z_b)\varphi(z_b) dz_b \nonumber \\[5pt]
   &= \int_0^\infty \varphi(z_b) dz_b \nonumber \\[5pt]
   &= \frac{1}{2}\left[1 + \text{erf}\left(\frac{\mu_b}{\sqrt{2\sigma^2_b}}\right)\right]
\end{align}
which then implies
\begin{align}
    \mathbb{E}[\text{ERUPT}[\hat{U}]] &\approx \sum_{b\in \mathcal{B}} \frac{p(b)}{2}\left[1 + \text{erf}\left(\frac{\mu_b}{\sqrt{2\sigma^2_b}}\right)\right](u(b) -c) \nonumber \\[5pt]
    &=: f_\alpha^\text{ERUPT}(\{n(b,t)\}; \theta^*)
\end{align}

\section{Connection with importance sampling}
\label{app:importance}

Here we briefly view our approach to task-specific experimentation from an alternative perspective. The different distributions of $n(x,t)$ values plotted in \Fig{intuition}(d) and \ref{fig:intuition}(e) might lead one to wonder whether there could be a connection between our method and importance sampling. 

A weak connection between our method and importance sampling exists when the target metric can be viewed as the error on the estimation of a particular integral. This is explicitly true for the squared ATE error, and it is also true for the AUQ but not for the MSE. (Up to a shift by a constant, the AUQ can be interpreted as the difference between the estimated integral $\text{AUQ}[\hat U]$ and true integral $\text{AUQ}[u]$.) In this section we will assume the context of the squared ATE error.

To make the connection, note that \Eq{ate} can be written as
\eqn{
\text{ATE} ~&=\!\!\! \int\displaylimits_{\mathcal{X},\,\mathcal{T},\,\mathcal{Y}} \!\!\! p(x) ~ p(y|x,t) ~ (-1)^{t+1} \, y ~ dy \, dt \, dx \nonumber \\
&=\!\!\! \int\displaylimits_{\mathcal{X},\,\mathcal{T},\,\mathcal{Y}} \!\!\! \frac{p(x) ~ p(y|x,t)}{h(x, t)} ~ (-1)^{t+1} \, y ~ h(x, t) ~ dy \, dt \, dx \nonumber \\
&\approx~ \sum_{i=1}^n \frac{p(X_i)}{n \cdot h(X_i, T_i)} \, (-1)^{T_i +1} \, Y_i|_{X_i, T_i} \label{eq:ate_importance}
}
where we have introduced an alternative sampling distribution $h(X, T)$ so that $h(X, T) \, p(Y|X, T)$ defines a joint probability distribution over $(\mathcal{X}, \mathcal{T}, \mathcal{Y})$. \Eq{ate_importance} provides an importance-sampling Monte Carlo estimate of the ATE that depends on $N$ independent samples $(X_i, T_i, Y_i)$ from this joint distribution. Note the similarity between \Eq{ate_importance} and $\text{ATE}[\hat U]$, i.e.~the result of using \Eq{bayesian_model} with $\alpha=\beta=0$ in order to give an estimate of \Eq{ate}. The only difference is that in $\text{ATE}[\hat U]$ the $n(x, t)$ counts are controlled manually, whereas in \Eq{ate_importance} the counts are stochastic, with expectation value $n \cdot h(x, t)$. We thus refer to \Eq{ate_importance} as $\text{ATE}[\hat U_h]$, as it can be interpreted as the ATE of an empirical uplift model defined using samples from $h(X, T)$.

The variance of $\text{ATE}[\hat U_h]$ is given by
\eqn{
&\frac{1}{n^2}\sum_{i=1}^n \var{\frac{p(X_i)}{h(X_i, T_i)} (-1)^{T_i +1}Y_i} \label{eq:ate_variance} \\[5pt]
& = \frac{1}{n}\e{\frac{p(X)^2}{h(X, T)^2} Y^2} -\frac{1}{n}\textnormal{ATE}^2\nonumber \\[5pt]
&= \int\displaylimits_{\mathcal{X},\,\mathcal{T}} \frac{p(x)^2 \, \e{Y^2|X=x,\,T=t}}{n \cdot h(x,t)} ~ dt \, dx ~-~ \frac{\text{ATE}^2}{n} \nonumber
}
It follows from the Cauchy-Schwarz inequality that the choice of $h(X, T)$ that minimises this variance is given by
\eqn{
h^*(X,T) = \frac{1}{Z} ~ p(x) \,\sqrt{\e{Y^2|X,T}}
}
where $Z$ is a constant normalisation factor. Sampling from $h^*(X, T)$ differs from our proposed approach to task-specific experimentation presented above, both because of the stochasticity in the number of samples chosen in each discrete bin (which is $n \cdot h(x,t)$ only in expectation) and because of the weaker connection with Thompson sampling (which is described precisely in \Sec{our_method} for the method of this paper but which does not carry over cleanly to the importance sampling setup). 

\Eq{ate_variance} can be useful though in testing whether a particular sampling distribution $h(X, T)$, e.g.~an ordinary RCT or an alternative proposal (such as uniform sampling in latent space), should be expected to perform well.

\end{document}